\documentclass[smallextended]{svjour3}       
\smartqed  

\usepackage{graphicx}
\usepackage{mathptmx}      
\usepackage{latexsym}
\usepackage{subfigure}
\usepackage{subfigmat}
\usepackage[colorlinks,linkcolor = blue,citecolor = blue]{hyperref}
\usepackage[cmex10]{amsmath}
\usepackage{amssymb}
\usepackage{epstopdf}
\usepackage{bm,upgreek}
\graphicspath{{figures/}}

\journalname{Journal of Scientific Computing}

\begin{document}

\title{
  Kinetic Simulation of Collisional Magnetized Plasmas with Semi-Implicit Time Integration
  \thanks{This work was performed under the auspices of the U.S. Department of Energy by
    Lawrence Livermore National Laboratory under contract DE-AC52-07NA27344.
    This material is based upon work supported by the U.S. Department of Energy, 
    Office of Science, Office of Advanced Scientific Computing Research, Applied 
    Mathematics program.
  }
}

\titlerunning{IMEX Mehods for Collisional Magnetized Plasmas}        

\author{Debojyoti~Ghosh \and
        Mikhail~A.~Dorf \and
        Milo~R.~Dorr \and
        Jeffrey~A.~F.~Hittinger
}


\institute{D. Ghosh \at
           Center for Applied Scientific Computing, Lawrence Livermore National Laboratory, Livermore, CA, 94550\\
           \email{ghosh5@llnl.gov} \\
           \and
           M. A. Dorf \at
           Physics Division, Lawrence Livermore National Laboratory, Livermore, CA, 94550\\
           \email{dorf1@llnl.gov}  \\
           \and
           M. R. Dorr \at
           Center for Applied Scientific Computing, Lawrence Livermore National Laboratory, Livermore, CA, 94550\\
           \email{dorr1@llnl.gov} \\
           \and
           J. A. F. Hittinger \at
           Center for Applied Scientific Computing, Lawrence Livermore National Laboratory, Livermore, CA, 94550\\
           \email{hittinger1@llnl.gov}
}

\date{Received: date / Accepted: date}

\maketitle

\vspace{-2.5in}
\hspace{-0.2in}
{\bf LLNL-JRNL-735522}
\hspace{0.2in}
\vspace{2.5in}

\begin{abstract}
Plasmas with varying collisionalities occur in many applications, such as tokamak edge regions, where the flows are characterized by significant variations in density and temperature. While a kinetic model is necessary for weakly-collisional high-temperature plasmas, high collisionality in colder regions render the equations numerically stiff due to disparate time scales. In this paper, we propose an implicit-explicit algorithm for such cases, where the collisional term is integrated implicitly in time, while the advective term is integrated explicitly in time, thus allowing time step sizes that are comparable to the advective time scales. This partitioning results in a more efficient algorithm than those using explicit time integrators, where the time step sizes are constrained by the stiff collisional time scales. We implement semi-implicit additive Runge-Kutta methods in COGENT, a finite-volume gyrokinetic code for mapped, multiblock grids and test the accuracy, convergence, and computational cost of these semi-implicit methods for test cases with highly-collisional plasmas.
\end{abstract}

\section{Introduction}
\label{intro}

\begin{figure}[t]
\begin{center}
\includegraphics[width=0.5\textwidth]{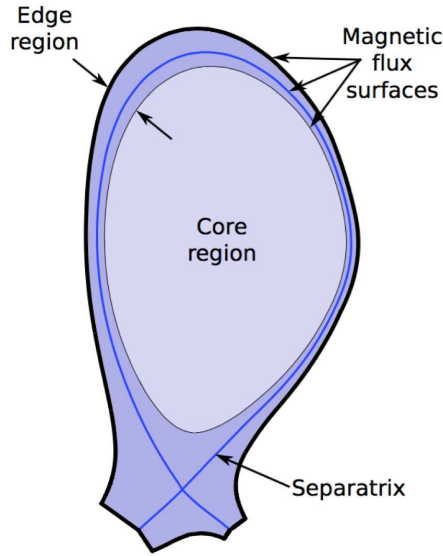}
\end{center}
\caption{Cross-section of a tokamak fusion reactor. The edge region is shaded in dark blue while the core region is shaded in light blue. The geometry is defined by magnetic flux surfaces. The separatrix (blue line) separates the open flux surfaces outside it from the closed flux surfaces inside it.}
\label{fig:tokamak}
\end{figure}

The purpose of this paper is to describe the application and performance of
a semi-implicit time integration algorithm for the solution of a system of
Vlasov-Fokker-Planck equations, motivated by the goal of simulating the
edge plasma region of tokamak fusion reactors. 
Plasma dynamics in the tokamak edge region is an unsteady multiscale phenomenon, characterized by a large range of spatial and temporal scales due to the density and temperature variations. Figure~\ref{fig:tokamak} shows the cross-section of a typical tokamak fusion reactor. The geometry is defined by the magnetic flux surfaces that contain the plasma, and the core and the edge regions are marked. Within the edge region, as the temperature decreases from the hot near-core region to the cold outer-edge region, there are three scale regimes. The hot and dense plasma in the inner edge region adjacent to the core plasma is weakly collisional, and the mean free paths of the particles are significantly larger than the density and temperature gradient length scales~\cite{cohenxu2008,dorf_etal_2013,dorf_etal_2014}. Near the separatrix that separates the closed magnetic field lines from the open ones, the plasma is moderately collisional, and the mean free paths are comparable to the density and temperature gradient scales. At the outer edge (near the material surfaces), the cold plasma is strongly collisional. In this region, the particle mean free paths are significantly smaller than the density and temperature length scales. 

Due to the weak collisionality near the core, a kinetic description with an appropriate collision model is required to model accurately the perturbations to the velocity distribution from the Maxwellian distribution. The Vlasov-Fokker-Planck (VFP) equation governs the evolution of the distribution function of each charged particle in the position and velocity space~\cite{thomasetal2012}. In the presence of a strong, externally-applied magnetic field, the ionized particles gyrate around the magnetic field lines. In the context of tokamaks, the radius of this gyromotion (gyroradius) is much smaller than the characteristic length scales. The gyrokinetic VFP equation represents the dynamics of the particles averaged over the gyromotion~\cite{lee1983,cohenxu2008}, i.e., it describes the motion of the guiding center of the particles. It is thus expressed in terms of the gyrocentric coordinates (parallel and perpendicular to the magnetic field lines) and does not contain phase-dependent terms. As a result, one velocity dimension and the fast time scale of the gyromotion is removed. The drift-kinetic model is the long wavelength limit of the gyrokinetic model, where turbulent length scales (comparable to the gyroradius) are neglected. We consider the drift-kinetic VFP equation in this paper.

Numerical algorithms to solve the gyrokinetic equation can be classified into three families. The Lagrangian particle-in-cell (PIC) approach~\cite{changku2008,heikkinenetal2001,heikkinenetal2006,chenetal2011,chenchacon2014,chenchacon2015} solves the equations of motion for ``superparticles" in the Lagrangian frame. The primary drawback is that the number of superparticles needed to control the numerical noise is very large~\cite{thomasetal2012,filbetetal2001}. Eulerian methods~\cite{idomuraetal2008_2,scott2006,xuetal2007,cohenxu2008,xuetal2008,xuetal2010} solve the VFP equation on a fixed phase space grid, where the phase space comprises the spatial position and velocity coordinates. Semi-Lagrangian approaches~\cite{gysela,rossmanithseal2011,crouseillesetal2009,qiuchristlieb2010,qiushu2011} follow characteristics, either tracing backward or forward in time, requiring interpolation of the solution from or to a fixed phase-space grid. In this paper, we adopt the Eulerian approach that allows the use of advanced numerical algorithms developed by the computational fluid dynamics (CFD) community. 

The gyrokinetic VFP equation is a parabolic partial differential equation (PDE) that comprises an advective Vlasov term and the Fokker-Planck collision term. The Vlasov term describes incompressible flow in the phase space~\cite{thomasetal2012,hahm1996}, i.e., the evolution of the distribution function by the gyrokinetic velocity of the particles and their acceleration parallel to the magnetic field. The Fokker-Planck term describes the collisional relaxation of the distribution function to the Maxwellian distribution~\cite{thomasetal2012,rosenbluth1957} and acts in the velocity space only. The difference in the time scales of these two terms varies significantly in the edge region due to the variation in the collisionality. The weakly and moderately collisional plasma at the inner edge and the separatrix, respectively, exhibits collisional time scales much smaller or comparable to the Vlasov time scales, while the strongly-collisional plasma at the outer edge exhibits collisional time scales much smaller than the Vlasov time scales. Consequently, the VFP equation exhibits disparate scales and is numerically stiff. Explicit time integration methods are inefficient because the time step size is constrained by the collisional time scale. Implicit time integration methods, on the other hand, are unconditionally linearly stable, but they require solutions to a linear or nonlinear system of equations. Linearly implicit~\cite{lemou2005,buetetal1997,lemou1998,larroche1993,larroche2003,lemou2004,casanovaetal1991} and nonlinearly implicit methods~\cite{larsenetal1985,epperlein1994,chaconetal2000_2,taitanoetal2015,taitanoetal2016} have been applied to the Fokker-Planck operator in isolation, and several implicit algorithms have also been proposed for the VFP equation~\cite{epperleinetal1988,mousseauknoll1997,kinghambell2004,thomasetal2009}. In our context, the accurate resolution of the unsteady dynamics requires time steps comparable to the Vlasov time scales, and thus a fully-implicit approach is inefficient.

Semi-implicit or implicit-explicit (IMEX) time integration methods~\cite{ascherruuthspiteri,pareschirusso,kennedycarpenter} allow the partitioning of the right-hand-side (RHS) into two parts: the stiff component integrated implicitly, comprising time scales faster than those of interest, and the nonstiff component integrated explicitly, comprising the slower scales. Such methods have been applied successfully to many multiscale applications~\cite{durranblossey,giraldorestellilauter2010,ghoshconstaSISC2016,jacobshesthaven2009,kumarmishra2012,kadiogluetal2010,salmietal2014,wolkeknoth2000}. In the context of the Vlasov equation, semi-implicit approaches have been proposed~\cite{chengknorr1976,candywaltz2003,idomuraetal2008_1,maeyamaetal2012} that partition the Vlasov term and integrate the parallel advection implicitly in time. 

In this paper, we propose a semi-implicit algorithm for the gyrokinetic VFP equation, where we integrate the Vlasov term explicitly while the stiff Fokker-Planck term is integrated implicitly to allow time steps that are comparable to the Vlasov time scales. The Fokker-Planck term represents a nonlinear advection-diffusion operator that can be expressed in two forms: the Landau form~\cite{landau1937} and the Rosenbluth form~\cite{rosenbluth1957}. Although the two forms are theoretically equivalent, they differ in their implications on the implicit numerical solution of this term. The Landau form expresses the advection and diffusion coefficients as direct integrals and is well-suited for conservative numerical methods~\cite{chang1970,kho1985,larsenetal1985,epperlein1994,lemou2005,buetcordier1998,berezinetal1987}. However, when integrated implicitly, the integral form results in a dense, nonlinear system of equations. Naive algorithms scale as $O\left(N^2\right)$, where $N$ is the number of velocity space grid points, whereas fast algorithms~\cite{filbet2002,buetetal1997,banksetal2016,brunneretal2010} scale as $O\left(N\log{N}\right)$. The Rosenbluth form, on the other hand, relates the advection and diffusion coefficients to the distribution function through Poisson equations for the Rosenbluth potentials in the velocity space. Thus, each evaluation of the Fokker-Planck term requires the solution to the Poisson equations, which scales as $O\left(N\right)$, assuming that an efficient Poisson solver is available. The primary difficulty is the need for solving the Poisson equations, defined on an infinite velocity space, on the truncated numerical domain, and several approaches have been proposed~\cite{james1977,bellicandy2012,mccoyetal1981,landremanernst2013,pataki2011,larroche1993,chaconetal2000_2}. An additional difficulty with the Rosenbluth form is the difficulty in enforcing mass, momentum, and energy conservation~\cite{chaconetal2000_1,taitanoetal2015}, and modifications are necessary to ensure energy and momentum are conserved to round-off errors. In this paper, we consider the Rosenbluth form of the Fokker-Planck term.

Our semi-implicit approach is implemented in COGENT~\cite{dorf_etal_2012,dorf_etal_2013,dorf_etal_2014}, a high-order finite-volume code that solves the gyrokinetic VFP equations on mapped, multiblock grids representing complex geometries. We implement multistage, conservative additive Runge-Kutta (ARK) methods~\cite{kennedycarpenter}, where the resulting nonlinear system of equations for the implicit stages are solved using the Jacobian-free Newton-Krylov approach~\cite{knollkeyes2004}. We investigate the performance of the ARK methods for test problems representative of the tokamak edge and compare it to that of the explicit Runge-Kutta (RK) methods for VFP problems. In particular, we verify the accuracy and convergence of the ARK methods and their computational efficiency with respect to the RK methods. Although the preconditioning that we have implemented reduces the cost of the implicit solve significantly, our approach is a preliminary effort, and the implementation of better preconditioning techniques will be addressed in the future.

The outline of the paper is as follows. Section~\ref{sec:govern} introduces the gyrokinetic VFP equations and the nondimensionalization used in our implementation. Section~\ref{sec:spatial} describes the high-order finite-volume method that COGENT uses to discretize the equations in space. The time integration methods are discussed in Section~\ref{sec:time}. The algorithm verification through two test cases is presented in Section~\ref{sec:results}, and Section~\ref{sec:summary} summarizes the contributions of this paper.

\section{Governing Equations}
\label{sec:govern}

The plasma dynamics at the tokamak edge are described by the full-$f$ gyrokinetic VFP equation~\cite{hahm1996,dorf_etal_2013}. In this paper, we consider a single species, and the axisymmetric governing equation is expressed in the nondimensional form as
\begin{align}
\frac {\partial B_\parallel^* f} {\partial t} + \nabla_{\bf R} \cdot \left( \dot{\bf R} B_\parallel^* f \right) + \frac {\partial} {\partial v_\parallel} \left( \dot{v}_\parallel B_\parallel^* f \right) = c_{f} \left( f \right),\label{eqn:gk}
\end{align}
where  $f \equiv f\left({\bf R},v_\parallel,\mu, t\right)$ is the distribution function defined on the phase space $\left({\bf R},v_\parallel,\mu\right)$. ${\bf R} \equiv \left(r,\theta\right)$ is the spatial gyrocenter position vector in the configuration space with $r$ as the radial coordinate and $\theta$ as the poloidal coordinate, $v_\parallel$ is the velocity parallel to the externally applied magnetic field ${\bf B}$, and $\mu = m v_\bot^2 / \left|{\bf B}\right|$ is the magnetic moment. 
The configuration space ${\bf R}$ is two-dimensional, and (\ref{eqn:gk}) is a four-dimensional (2D--2V) PDE. 
$\nabla_{\bf R}$ denotes the divergence operator in the configuration space. 
In the long wavelength limit, the gyrokinetic model reduces to the drift-kinetic model, and the Vlasov velocity and parallel acceleration are given by
\begin{subequations}\label{eqn:vlasov_term}
\begin{align}
\dot{\bf R} &\equiv \dot{\bf R} \left({\bf R}, v_\parallel, \mu, t \right) = \frac {1} {B^*_\parallel} \left[ v_\parallel {\bf B}^* + \frac {\lambda_a} {Z} \hat{\bf b} \times \left( Z {\bf E} + \frac{\mu}{2} \nabla_{\bf R} B \right) \right], \\
\dot{v}_\parallel &\equiv \dot{v}_\parallel \left( {\bf R}, v_\parallel, \mu, t \right) = - \frac {1}{m B^*_\parallel} {\bf B}^* \cdot \left( Z {\bf E} + \frac{\mu}{2} \nabla_{\bf R} B \right),
\end{align}
\end{subequations}
where $m$ and $Z$ are the mass and ionization state, respectively, $\lambda_a$ is the Larmor number (ratio of the gyroradius to the characteristic length scale), $\hat{\bf b} = {\bf B}/\left|{\bf B}\right|$ and $B = \left|{\bf B}\right|$ are the unit vector along the magnetic field and magnitude of the magnetic field, and
\begin{align}
{\bf B}^* \equiv {\bf B}^* \left( {\bf R}, v_\parallel \right) = {\bf B} + \lambda_a \frac {m v_\parallel} {Z} \nabla \times \hat{\bf b}.
\end{align}
$B_{\parallel}^* = {\bf B}^* \cdot \hat{\bf b}$ is the Jacobian of the transformation from the lab frame to the gyrocentric coordinates. In this paper, we consider cases with uniform magnetic field, and thus ${\bf B}^* \equiv {\bf B}$. The electric field is ${\bf E} = -\nabla_{\bf R} \phi$, where $\phi$ is the electrostatic potential. In this paper, we consider cases where $\phi = \phi\left({\bf R}\right)$ is specified; however, COGENT generally computes a self-consistent electrostatic potential from the species charge densities~\cite{dorf_etal_2012,dorf_etal_2013} where a Boltzman or vorticity model is assumed for the electrons.

The Rosenbluth form of the Fokker-Planck collision term for a single species is expressed as~\cite{dorf_etal_2014}
\begin{subequations}\label{eqn:fp}
\begin{align}
c \left( f \right) &= \nu_c \nabla_{\left(v_\parallel,\mu\right)} \cdot \overrightarrow{\Gamma}, \label{eqn:fp_1}\\
\overrightarrow{\Gamma} &= \overrightarrow{\sigma} f + \overleftrightarrow{{\kappa}} \nabla_{\left(v_\parallel,\mu\right)} f, \label{eqn:fp_flux_1}
\end{align}
\end{subequations}
where $\nu_c$ is the collision frequency, and the coefficients are defined as follows:
\begin{align}
\overrightarrow{\sigma} = \left[\begin{array}{c} \sigma_{v_\parallel} \\ \sigma_{\mu}  \end{array}\right],
\overleftrightarrow{{\kappa}} = \left[ \begin{array}{cc} \kappa_{v_\parallel v_\parallel} & \kappa_{v_\parallel \mu} \\ \kappa_{\mu v_\parallel} & \kappa_{\mu\mu} \end{array}\right],
\end{align}
where
\begin{align}
&\sigma_{v_\parallel} = \frac {\partial \varphi} {\partial v_\parallel},\ \ \ \sigma_{\mu} = 4\mu\left(\frac{m}{B}\right)\frac {\partial \varphi} {\partial \mu},\ 
\kappa_{v_\parallel v_\parallel} = - \frac {\partial^2 \varrho} {\partial v_\parallel^2}, \nonumber\\
&\kappa_{v_\parallel \mu} = \kappa_{\mu v_\parallel} = - 4\mu\left(\frac{m}{B}\right) \frac {\partial^2 \varrho} {\partial v_\parallel \partial \mu},\ 
\kappa_{\mu\mu} = -8\mu \left( \frac {m} {B} \right)^2 \left[ 2\mu \frac {\partial^2 \varrho } {\partial \mu^2} + \frac {\partial \varrho} {\partial \mu} \right].\label{eqn:fpcoeffs}
\end{align}
The operator $\nabla_{\left(v_\parallel,\mu\right)}$ denotes the gradient operator in the velocity space $\left(\partial/\partial v_\parallel, \partial/\partial\mu\right)$.
The Rosenbluth potentials, $\varphi$ and $\varrho$, are related to the distribution function through the following Poisson equations in the two-dimensional velocity space $\left(v_\parallel,\mu\right)$:
\begin{subequations}\label{eqn:rosenbluth}
\begin{align}
\frac {\partial^2 \varphi} {\partial v_\parallel^2} + \frac {m}{B} \frac {\partial} {\partial \mu} \left( 2\mu \frac{\partial \varphi} {\partial \mu} \right) &= f,\label{eqn:rosenbluth1} \\
\frac {\partial^2 \varrho} {\partial v_\parallel^2} + \frac {m}{B} \frac {\partial} {\partial \mu} \left( 2\mu \frac{\partial \varrho} {\partial \mu} \right) &= \varphi.\label{eqn:rosenbluth2}
\end{align}
\end{subequations}
The Maxwellian distribution function, defined as
\begin{align}
f_{\rm M} \left({\bf R},v_\parallel, \mu\right) = \frac{n\left({\bf R}\right)}{\sqrt{\pi}} \left( \frac{m}{2T\left({\bf R}\right)} \right)^{\frac{3}{2}}  \exp\left(- \frac{m\left\{v_\parallel-\overline{v_\parallel}\left({\bf R}\right)\right\}^2+\mu B\left({\bf R}\right)}{2T\left({\bf R}\right)}\right),
\end{align}
is an exact solution of the Fokker-Planck collision operator ($c\left(f_{\rm M}\right) = 0$),
where the temperature $T\left({\bf R}\right)$ is given by
\begin{align}
T\left({\bf R}\right) &= \frac{1}{n\left({\bf R}\right)} \int\int \frac{1}{3} \left[ \left(v_\parallel - \overline{v_\parallel}\left({\bf R}\right)\right)^2 + \mu B \right] f\left({\bf R},v_\parallel,\mu\right) B^*_\parallel\left({\bf R},v_\parallel\right) dv_\parallel d\mu
\end{align}
and
\begin{align}
n\left({\bf R}\right) &= \frac{1}{m} \int\int f\left({\bf R},v_\parallel,\mu\right) B^*_\parallel\left({\bf R},v_\parallel\right) dv_\parallel d\mu,\\
\overline{v_\parallel}\left({\bf R}\right) &= \frac{1}{m\,n\left({\bf R}\right)} \int\int v_\parallel f\left({\bf R},v_\parallel,\mu\right) B^*_\parallel\left({\bf R},v_\parallel\right) dv_\parallel d\mu
\end{align}
are the number density and average parallel velocity, respectively. 

The equations in the preceding discussion involve non-dimensional variables. The non-dimensionalization is derived by
\begin{align}
&t = \frac{\tilde{t}}{t_{\rm ref}},\ \ 
v_\parallel = \frac{\tilde{v_\parallel}}{v_{\rm ref}},\ \ 
n = \frac{\tilde{n}}{n_{\rm ref}},\ \ 
m = \frac{\tilde{m}}{m_{\rm ref}},\ \ 
f = \frac{\tilde{f}}{f_{\rm ref}},\ \ 
T = \frac{\tilde{T}}{T_{\rm ref}},\nonumber\\
&B = \frac{\tilde{B}}{B_{\rm ref}},\ \ 
\phi = \frac{\tilde{\phi}}{\phi_{\rm ref}},\ \ 
\mu = \frac{\tilde{\mu}}{\mu_{\rm ref}},\ \ \ \ 
x,y,L = \frac{\tilde{x},\tilde{y},\tilde{L}}{L_{\rm ref}},\nonumber
\end{align}
where $\tilde{\left(\cdot\right)}$ is the physical (dimensional) quantity and $\left(\cdot\right)_{\rm ref}$ are the reference quantities. The Larmor number and the collision frequency are expressed as
\begin{align}
\lambda_a = \frac{m_{\rm ref} v_{\rm ref}}{e L_{\rm ref} B_{\rm ref}},\ \ \ \ \ 
\nu_c = \frac {f_{\rm ref} t_{\rm ref} e^4} {m_{\rm ref}^2} \Lambda_c \left(\frac{4\pi Z^2}{m}\right)^2,\nonumber
\end{align}
respectively, where $\Lambda_c$ is the Coulomb logarithm. The primitive reference parameters are $n_{\rm ref}$ (number density), $T_{\rm ref}$ (temperature), $L_{\rm ref}$ (length), $m_{\rm ref}$ (mass), and $B_{\rm ref}$ (magnetic field); the derived reference quantities are
\begin{align}
& v_{\rm ref} = \sqrt{\frac{T_{\rm ref}}{m_{\rm ref}}}\ {(\rm thermal\ speed)},\ \ 
t_{\rm ref} = \frac{L_{\rm ref}}{v_{\rm ref}}\ {(\rm transit\ time)}, \nonumber\\
& \mu_{\rm ref} = \frac{T_{\rm ref}}{2B_{\rm ref}}\ {(\rm magnetic\ moment)},\ \ 
f_{\rm ref} = \frac{n_{\rm ref}}{\pi v_{\rm ref}^3}\ {(\rm distribution\ function)},\nonumber \\
& \phi_{\rm ref} = \frac{T_{\rm ref}}{e}\ {(\rm potential)};\nonumber
\end{align}
and $e$ is the elementary charge.

\section{Spatial Discretization}
\label{sec:spatial}

Equation~(\ref{eqn:gk}) is a parabolic PDE, which we discretize in space using a fourth-order finite-volume method on a mapped grid~\cite{collelaetal2011,mccorquodaleetal2015}. The computational domain is defined as a four-dimensional hypercube of unit length in each dimension,
\begin{align}
\Omega = \left\{ \bm{\xi} : 0 \le \xi_d \le 1, 1 \le d \le D \right\};\ \ \xi_d = \bm{\xi} \cdot {\bf e}_d,
\end{align}
partitioned by a uniform grid with a computational cell defined as
\begin{align}
\upomega_{\bf i} = \prod_{d=1}^{D} \left[ \left({\bf i} - \frac{1}{2}{\bf e}_d \right)h, \left({\bf i} + \frac{1}{2}{\bf e}_d \right)h \right],
\end{align}
where $D=4$ is the number of spatial dimensions, ${\bf e}_d$ is the unit vector along dimension $d$, and ${\bf i}$ is an integer vector representing a four-dimensional grid index. The transformation ${\bf x} = {\bf X}\left(\bm{\xi}\right)$ defines the mapping between a vector in the physical space ${\bf x}\equiv \left({\bf R},v_\parallel, \mu \right)$ and a vector in the computational space $\bm{\xi}$. Equation~(\ref{eqn:gk}) is integrated over a physical cell ${\bf X}\left(\upomega_{\bf i}\right)$ to yield the integral form
\begin{align}
\frac {\partial} {\partial t} \left( \int\displaylimits_{{\bf X}\left(\upomega_{\bf i}\right)} {B_\parallel^*f} d{\bf x} \right)  = \int\displaylimits_{{\bf X}\left(\upomega_{\bf i}\right)} \mathcal{V}\left({B_\parallel^*f}\right) d{\bf x} + \int\displaylimits_{{\bf X}\left(\upomega_{\bf i}\right)} \mathcal{C}\left({f}\right) d{\bf x},\label{eqn:integral_form}
\end{align}
where the Vlasov and collision terms are
\begin{align}
\mathcal{V}\left({B_\parallel^*f}\right) \equiv - \left[ \nabla_{\bf R} \cdot \left( \dot{\bf R} {B_\parallel^*f} \right) + \frac {\partial} {\partial v_\parallel} \left( \dot{v}_\parallel {B_\parallel^*f} \right) \right],\ \ 
\mathcal{C}\left({f}\right) = c_f \left( {f} \right).
\end{align}
Defining the computational cell-averaged solution as
\begin{align}
\bar{f}_{\bf i} = \frac {1} {\underline{\upomega_{\bf i}}} \int\displaylimits_{\upomega_{\bf i}} {B_\parallel^*f} J d\bm{\xi};\ \ \ J \equiv \left| \frac {\partial {\bf x}} {\partial \bm{\xi}}  \right|,
\end{align}
where $\underline{\upomega_{\bf i}}$ is the volume of the computational cell $\upomega_{\bf i}$, the physical cell-averaged solution is
\begin{align}
\breve{f}_{\bf i} = \frac{1}{\underline{{\bf X}\left(\upomega_{\bf i}\right)}} \int\displaylimits_{{\bf X}\left(\upomega_{\bf i}\right)} {B_\parallel^*f} d{\bf x} = \left( \int\displaylimits_{{\bf V}_{\bf i}} J d\bm{\xi} \right)^{-1} \int\displaylimits_{\upomega_{\bf i}} {B_\parallel^*f} J d\bm{\xi} = \bar{J}_{\bf i}^{-1} \bar{f}_{\bf i},\ 
{\rm where}\ 
\bar{J}_{\bf i} = \frac{1}{\underline{\upomega_{\bf i}}} \int\displaylimits_{\upomega_{\bf i}} J d\bm{\xi}.
\end{align}
The Vlasov and collision terms in (\ref{eqn:integral_form}) can be written in the divergence form as follows:
\begin{subequations}
\begin{align}
\mathcal{V}\left({B_\parallel^*f}\right) &= \nabla_{\bf x} \cdot {\bf V}\left({B_\parallel^*f}\right);\ \ {\bf V}\left({B_\parallel^*f}\right) = -\left[\begin{array}{ccccccc} \left(\dot{\bf R}\cdot \hat{\bf r}\right){B_\parallel^*f}, & &  \left(\dot{\bf R}\cdot \hat{\boldsymbol\theta}\right){B_\parallel^*f}, & & \dot{v}_\parallel{B_\parallel^*f}, & & 0 \end{array}\right]^{\rm T}, \\
\mathcal{C}\left({f}\right) &= \nabla_{\bf x} \cdot {\bf C}\left({f}\right);\ \ {\bf C}\left({f}\right) = \nu_c \left[\begin{array}{ccccccc} 0, & & 0, & & \overrightarrow{\Gamma}\cdot\hat{{\bf v}_\parallel}, & & \overrightarrow{\Gamma}\cdot\hat{\boldsymbol\mu} \end{array}\right]^{\rm T},
\end{align}
\end{subequations}
where $\hat{\bf r},\hat{\boldsymbol\theta},\hat{{\bf v}_\parallel},\hat{\boldsymbol\mu}$ are the unit vectors along the $r,\theta,v_\parallel,\mu$ coordinates, respectively, and $\nabla_{\bf x}$ denotes the divergence operator in the physical space. Using the divergence theorem, (\ref{eqn:integral_form}) is expressed as
\begin{align}
\frac {\partial \bar{f}_{\bf i}} {\partial t} &= \frac{1}{\underline{\upomega_{\bf i}}} \left[ \int\displaylimits_{\partial{\bf X}\left(\upomega_{\bf i}\right)} {\bf V}\left({B_\parallel^*f}\right) \cdot \hat{\bf n} d{\bf s} + \int\displaylimits_{\partial{\bf X}\left(\upomega_{\bf i}\right)} {\bf C}\left({f}\right) \cdot \hat{\bf n} d{\bf s} \right] \nonumber\\
&= \frac{1}{\underline{\upomega_{\bf i}}} \sum_{d=1}^{D} \left[ \int\displaylimits_{A_{ {\bf i}+\frac{1}{2}{\bf e}_d} }  {\bf N}^{\rm T}\left({\bf V}+{\bf C}\right) dA_{\bm{\xi}} - \int\displaylimits_{A_{ {\bf i}-\frac{1}{2}{\bf e}_d} }  {\bf N}^{\rm T}\left({\bf V}+{\bf C}\right) dA_{\bm{\xi}} \right],\label{eqn:integral_form2}
\end{align}
where $\hat{\bf n}$ is the outward normal for $\partial {\bf X}\left(\upomega_{\bf i}\right)$, ${\bf N} = J\nabla_{\bf x}\bm{\xi}$, and $A_{ {\bf i}\pm\frac{1}{2}{\bf e}_d}$ denote the faces along dimension $d$ of the cell $\upomega_{\bf i}$. The face-averaged Vlasov and collision fluxes on a given face are defined as
\begin{subequations}\label{eqn:faceavgflux}
\begin{align}
\left\langle\hat{\bf V}_{{\bf i}\pm\frac{1}{2}{\bf e}_d}\right\rangle = \frac{1}{\left|A_{ {\bf i}\pm\frac{1}{2}{\bf e}_d}\right|} \int\displaylimits_{A_{ {\bf i}\pm\frac{1}{2}{\bf e}_d} }  {\bf N}^{\rm T}{\bf V} dA_{\bm{\xi}};\ \  \left\langle \hat{V}_{{\bf i}\pm\frac{1}{2}{\bf e}_d} \right\rangle = \left\langle \hat{\bf V}_{{\bf i}\pm\frac{1}{2}{\bf e}_d}\right\rangle  \cdot e_d,\\
\left\langle\hat{\bf C}_{{\bf i}\pm\frac{1}{2}{\bf e}_d}\right\rangle = \frac{1}{\left|A_{ {\bf i}\pm\frac{1}{2}{\bf e}_d}\right|} \int\displaylimits_{A_{ {\bf i}\pm\frac{1}{2}{\bf e}_d} }  {\bf N}^{\rm T}{\bf C} dA_{\bm{\xi}};\ \  \left\langle \hat{C}_{{\bf i}\pm\frac{1}{2}{\bf e}_d} \right\rangle = \left\langle \hat{\bf C}_{{\bf i}\pm\frac{1}{2}{\bf e}_d}\right\rangle  \cdot e_d.
\end{align}
\end{subequations}
The computational cell is a hypercube with length $h$ in each dimension with the volume as $\underline{\upomega_{\bf i}} = h^D$, and the area of each face as $\left|A^d_{{\bf i}\pm\frac{1}{2}{\bf e}_d}\right| = h^{D-1},\ \forall d$. Thus, (\ref{eqn:integral_form2}) can be written as
\begin{align}
\frac {\partial \bar{f}_{\bf i}} {\partial t} = \frac{1}{h} \sum_{d=1}^{D} \left[ \left(\left\langle\hat{V}_{{\bf i}+\frac{1}{2}{\bf e}_d}\right\rangle - \left\langle\hat{V}_{{\bf i}-\frac{1}{2}{\bf e}_d}\right\rangle\right) + \left(\left\langle\hat{C}_{{\bf i}+\frac{1}{2}{\bf e}_d}\right\rangle - \left\langle\hat{C}_{{\bf i}-\frac{1}{2}{\bf e}_d}\right\rangle\right) \right].\label{eqn:fv}
\end{align}
Defining the solution vector as consisting of the cell-averaged distribution function at all the computational cells in the grid, (\ref{eqn:fv}) is expressed for the entire domain as a system of ordinary differential equations (ODEs) in time,
\begin{subequations}\label{eqn:semidisc}
\begin{align}
\frac {d \bar{\bf f}} {dt} &= \hat{\mathcal{V}}\left(\bar{\bf f}\right) + \hat{\mathcal{C}}\left(\bar{\bf f}\right);\ \ 
\bar{\bf f} = \left[\bar{f}_{\bf i} \right],\ \ {\bf i} \in \left\{ {\bf j} : \omega_{\bf j} \in \Omega \right\};\label{eqn:gk_semidisc}\\
\hat{\mathcal{V}}\left(\bar{f}_{\bf i}\right) &=  \frac{1}{h} \sum_{d=1}^{D} \left(\left\langle\hat{V}_{{\bf i}+\frac{1}{2}{\bf e}_d}\right\rangle - \left\langle\hat{V}_{{\bf i}-\frac{1}{2}{\bf e}_d}\right\rangle\right) = \mathcal{V}\left({B_\parallel^*f}\right) + \mathcal{O}\left( \Delta{\bf R}^p, \Delta v_\parallel^p, \Delta \mu^p \right),\label{eqn:semidisc_vlasov}\\
\hat{\mathcal{C}}\left(\bar{f}_{\bf i}\right) &=  \frac{1}{h} \sum_{d=1}^{D} \left(\left\langle\hat{C}_{{\bf i}+\frac{1}{2}{\bf e}_d}\right\rangle - \left\langle\hat{C}_{{\bf i}-\frac{1}{2}{\bf e}_d}\right\rangle\right) = \mathcal{C}\left({f}\right) + \mathcal{O}\left( \Delta{\bf R}^q, \Delta v_\parallel^q, \Delta \mu^q \right),\label{eqn:semidisc_collision}
\end{align}
\end{subequations}
where $p,q$ are the orders of the schemes used to discretize the Vlasov and collision terms, respectively.

Equation~(\ref{eqn:semidisc}) requires the computation of the face-averaged Vlasov and collisional fluxes defined by (\ref{eqn:faceavgflux}), for which we use the discretization described in~\cite{collelaetal2011}. The following relationships between the cell-centered and face-centered quantities and cell-averaged and face-averaged quantities to fourth order will be used in the subsequent discussion:
\begin{subequations}\label{eqn:conversions}
\begin{align}
\left\langle u \right\rangle_{{\bf i}\pm\frac{1}{2}{\bf e}_d} &= u_{{\bf i}\pm\frac{1}{2}{\bf e}_d} + \frac{h^2}{24} \sum_{\substack{d'=1\\d'\ne d}}^D \left. \frac{\partial^2 u}{\partial \xi_{d'}^2} \right|_{{\bf i}\pm\frac{1}{2}{\bf e}_d} + \mathcal{O}\left(h^4\right), \label{eqn:faceconv}\\
\bar{u}_{\bf i} &= u_{\bf i} + \frac{h^2}{24} \sum_{d=1}^D \left. \frac{\partial^2 u}{\partial \xi_d^2} \right|_{\bm{\xi}_{\bf i}} + \mathcal{O}\left(h^4\right),\label{eqn:volconv}
\end{align}
\end{subequations}
for an arbitrary variable $u$, where $\left\langle u \right\rangle_{{\bf i}\pm\frac{1}{2}{\bf e}_d}$ denotes the face-averaged value, $\bar{u}_{\bf i}$ denotes the cell-averaged value, and $u_{{\bf i}\pm\frac{1}{2}{\bf e}_d}$ and $ u_{\bf i}$ denote the face-centered and cell-centered values, respectively. We note that it is sufficient to compute the second derivatives in the RHS of (\ref{eqn:conversions}) to second order using centered finite-differences since they are multiplied by $h^2$. The face-averaged fluxes in (\ref{eqn:semidisc}) are computed to fourth order as
\begin{subequations}
\begin{align}
\hat{V}^d_{{\bf i}\pm\frac{1}{2}{\bf e}_d} = &\sum_{s=1}^{D} \left\langle N_d^s \right\rangle_{{\bf i}\pm\frac{1}{2}{\bf e}_d}\left\langle V^s \right\rangle_{{\bf i}\pm\frac{1}{2}{\bf e}_d} \nonumber \\& + h^2 \sum_{s=1}^{D} \left\{ {\bf G}_0^{\bot,d} \left\langle N_d^s \right\rangle_{{\bf i}\pm\frac{1}{2}{\bf e}_d} \right\} \cdot \left\{ {\bf G}_0^{\bot,d} \left\langle V^s \right\rangle_{{\bf i}\pm\frac{1}{2}{\bf e}_d} \right\} + \mathcal{O}\left(h^4\right),\label{eqn:vlasovflux}\\
\hat{C}^d_{{\bf i}\pm\frac{1}{2}{\bf e}_d} = &\sum_{s=1}^{D} \left\langle N_d^s \right\rangle_{{\bf i}\pm\frac{1}{2}{\bf e}_d}\left\langle C^s \right\rangle_{{\bf i}\pm\frac{1}{2}{\bf e}_d} \nonumber \\& + h^2 \sum_{s=1}^{D} \left\{ {\bf G}_0^{\bot,d} \left\langle N_d^s \right\rangle_{{\bf i}\pm\frac{1}{2}{\bf e}_d} \right\} \cdot \left\{ {\bf G}_0^{\bot,d} \left\langle C^s \right\rangle_{{\bf i}\pm\frac{1}{2}{\bf e}_d} \right\} + \mathcal{O}\left(h^4\right),\label{eqn:collflux}
\end{align}
\end{subequations}
where ${\bf G}_0^{\bot,d} \approx \nabla_{\bm{\xi}} - {\bf e}^d\frac{\partial}{\partial \xi_d}$ and $\left\langle N_d^s \right\rangle_{{\bf i}\pm\frac{1}{2}{\bf e}_d}; s=1,\cdots,D$ are the column vectors of the face-averaged metric quantities~\cite{collelaetal2011}.

Equation~(\ref{eqn:vlasovflux}) requires computing $\left\langle V^s \right\rangle_{{\bf i}\pm\frac{1}{2}{\bf e}_d}, s=1,\cdots,D$. Writing each component of four-dimensional Vlasov term as an advection operator,
\begin{align}
V^s = a^s {f},\ \ s=1,\cdots,D;\ \ \ \ 
a^s = \left\{\begin{array}{lcl} 
\left(\dot{\bf R}\cdot \hat{r}\right) & , & s = d_r\\
\left(\dot{\bf R}\cdot \hat{\theta}\right) & , & s = d_\theta \\
\dot{v}_\parallel & , & s = d_{v_\parallel} \\
0 & , & s = d_\mu
\end{array}\right.,
\end{align}
where $d_r, d_\theta, d_{v_\parallel}, d_\mu$ are the spatial dimensions corresponding to $r,\theta,v_\parallel,\mu$ coordinates, respectively, the face-averaged flux $\left\langle V^s \right\rangle_{{\bf i}\pm\frac{1}{2}{\bf e}_d}$ is computed to fourth order from the discrete convolution as
\begin{align}
\left\langle V^s \right\rangle_{{\bf i}\pm\frac{1}{2}{\bf e}_d} = \left\langle a^s\right\rangle_{{\bf i}\pm\frac{1}{2}{\bf e}_d} \left\langle \bar{\bar{f}} \right\rangle_{{\bf i}\pm\frac{1}{2}{\bf e}_d} + \frac{h^2}{12} \sum_{d' \ne d} \left( \frac{\partial a^s}{\partial \xi_{d'}} \frac{\partial \bar{\bar{f}}}{\partial \xi_{d'}} \right)_{\bm{\xi}_{{\bf i}\pm\frac{1}{2}{\bf e}_d}} + \mathcal{O}\left(h^4\right).\label{eqn:fluxavg}
\end{align}
The face-averaged advective coefficients $\left\langle a^s\right\rangle_{{\bf i}\pm\frac{1}{2}{\bf e}_d}$ are computed by evaluating $\dot{\bf R}$ and $\dot{v}_\parallel$ at the face centers using (\ref{eqn:vlasov_term}) and transforming them to face-averaged quantities using (\ref{eqn:faceconv}). The cell-averaged distribution function in the computational space $\bar{\bar{f}}_{\bf i}$ is defined and related to $\bar{f}_{\bf i}$ as follows:
\begin{align}
\bar{\bar{f}}_{\bf i} = \frac{1}{V_{\bf i}} \int\displaylimits_{V_{\bf i}} {B_\parallel^*f} d{\bm{\xi}} \ \ \ \Rightarrow \bar{\bar{f}}_{\bf i} = J^{-1}_{\bf i} \left[ \bar{f}_{\bf i} - \frac{h^2}{12} \nabla_{\bf\xi} \bar{f} \cdot \nabla_{\bf\xi}J \right] + \mathcal{O}\left(h^4\right).\label{eqn:fbarbar}
\end{align}
The derivatives $\partial/\partial\xi_{d'}$ in (\ref{eqn:fluxavg}) and $\nabla_\xi$ in (\ref{eqn:fbarbar}) are computed using second-order central finite-differences since they are multiplied by $h^2$. Finally, the face-averaged distribution function in the computational space $ \left\langle \bar{\bar{f}} \right\rangle_{{\bf i}\pm\frac{1}{2}{\bf e}_d}$ is computed from cell-averaged values using the fifth-order weighted essentially nonoscillatory (WENO) scheme~\cite{jiangshu} with upwinding based on the sign of $\left\langle a^s\right\rangle_{{\bf i}\pm\frac{1}{2}{\bf e}_d}$. Thus, the algorithm to compute $\hat{\mathcal{V}}\left(\bar{\bf f}\right)$ in (\ref{eqn:gk_semidisc}) is fourth-order accurate, and $p=4$ in (\ref{eqn:semidisc_vlasov}).

Equation~(\ref{eqn:collflux}) requires the computation of the face-averaged collision flux $\left\langle C^s \right\rangle_{{\bf i}\pm\frac{1}{2}{\bf e}_d}$. We note that the velocity space grid is Cartesian in our drift-kinetic model and independent of the configuration space grid; thus, $\left\langle C^s \right\rangle_{{\bf i}\pm\frac{1}{2}{\bf e}_d} = 0$ if $s \ne d$, and (\ref{eqn:collflux}) only needs $\left\langle C^d \right\rangle_{{\bf i}\pm\frac{1}{2}{\bf e}_d}$. The face-averaged flux is obtained from the face-centered collision flux 
\begin{align}
C^d_{{\bf i}\pm\frac{1}{2}{\bf e}_d} = \left\{\begin{array}{lcl} \Gamma^{v_\parallel}_{{{\bf i}\pm\frac{1}{2}{\bf e}_{v_\parallel}}} & , & d = d_{v_\parallel} \\ \Gamma^{\mu}_{{{\bf i}\pm\frac{1}{2}{\bf e}_\mu}} & , & d = d_\mu \\ 0 & , & {\rm otherwise} \end{array}\right.; \ \ \ 
\left.\begin{array}{l}
\Gamma^{v_\parallel}_{{{\bf i}\pm\frac{1}{2}{\bf e}_{v_\parallel}}}=\overrightarrow{\Gamma}_{{{\bf i}\pm\frac{1}{2}{\bf e}_{v_\parallel}}}\cdot\hat{{\bf v}_\parallel},\\ 
\Gamma^{\mu}_{{{\bf i}\pm\frac{1}{2}{\bf e}_\mu}} = \overrightarrow{\Gamma}_{{\bf i}\pm\frac{1}{2}{\bf e}_\mu}\cdot\hat{\boldsymbol\mu} 
\end{array}
\right.
\end{align} 
using (\ref{eqn:faceconv}). Equation~(\ref{eqn:fp_1}) involves derivatives in the velocity space $\left(v_\parallel,\mu\right)$ only, and to simplify the subsequent discussion, two-dimensional grid indices $\left(k,l\right)$ will be used, such that
\begin{align}
u_{k+p,l+q} \equiv u_{{\bf i} + p {\bf e}_{{v_\parallel}} + q {\bf e}_{\mu} };\ \ k = {\bf i}\cdot{\bf e}_{{v_\parallel}},\ \ l = {\bf i}\cdot{\bf e}_{\mu} 
\end{align}
for an arbitrary grid variable $u$ and $p,q \in \mathbb{Z}$. The computation of $\Gamma^{v_\parallel}_{{{\bf i}\pm\frac{1}{2}{\bf e}_{{v_\parallel}}}}$ is described in the following paragraphs and can be extended trivially to computing $\Gamma^{\mu}_{{{\bf i}\pm\frac{1}{2}{\bf e}_{\mu}}}$.

Evaluating (\ref{eqn:fp_flux_1}) at the face ${\bf i}+\frac{1}{2}{\bf e}_{v_\parallel}$, the Fokker-Planck collision flux along the $v_\parallel$ dimension is given by
\begin{align}
\Gamma^{v_\parallel}_{k+\frac{1}{2},l} \equiv \Gamma^{v_\parallel}_{{{\bf i}+\frac{1}{2}{\bf e}_{{v_\parallel}}}} = \left[ \sigma_{v_\parallel} f + \kappa_{v_\parallel v_\parallel} \frac{\partial f}{\partial v_\parallel} + \kappa_{v_\parallel \mu} \frac{\partial f}{\partial \mu} \right]_{k+\frac{1}{2},l}.\label{eqn:fp_flux}
\end{align}
The face-centered advective term is computed as
\begin{align}
\left[\sigma_{v_\parallel} f\right]_{k+\frac{1}{2},l} = \sigma_{v_\parallel,k+\frac{1}{2},l} f_{k+\frac{1}{2},l},\label{eqn:fp_adv_1}
\end{align}
where the distribution function at the interface $f_{k+\frac{1}{2},l}$ is computed using the fifth-order upwind interpolation method based on the sign of $\sigma_{v_\parallel,k+\frac{1}{2},l}$, i.e.,
\begin{subequations}
\begin{align}
\sigma_{v_\parallel,k+\frac{1}{2},l} &< 0, \nonumber\\
f_{k+\frac{1}{2},l} &= \frac{1}{30}f_{k-2,l} - \frac{13}{60}f_{k-1,l} + \frac{47}{60}f_{k,l} + \frac{27}{60}f_{k+1,l} - \frac{1}{20}f_{k+2,l}, \\ 
\sigma_{v_\parallel,k+\frac{1}{2},l} &\geq 0, \nonumber\\
f_{k+\frac{1}{2},l} &= \frac{1}{30}f_{k+3,l} - \frac{13}{60}f_{k+2,l} + \frac{47}{60}f_{k+1,l} + \frac{27}{60}f_{k,l} - \frac{1}{20}f_{k-1,l}.
\end{align}
\end{subequations}
Note that the advective term is on the RHS of (\ref{eqn:gk}), and a negative value of $\sigma_{v_\parallel,k+\frac{1}{2},l}$ implies a right-moving wave. The cell-centered values of the distribution function $f_{k,l}$ are obtained from its cell-averaged values using (\ref{eqn:volconv}). The diffusion terms are computed as follows:
\begin{align}
\left[ \kappa_{v_\parallel v_\parallel} \frac{\partial f}{\partial v_\parallel} \right]_{k+\frac{1}{2},l} = \kappa_{v_\parallel v_\parallel,{k+\frac{1}{2},l}} \left.\frac{\partial f}{\partial v_\parallel}\right|_{k+\frac{1}{2},l},\label{eqn:fp_diff_1}
\end{align}
and
\begin{align}
\left[ \kappa_{v_\parallel \mu} \frac{\partial f}{\partial \mu} \right]_{k+\frac{1}{2},l} = \frac{1}{2} \left( \kappa_{v_\parallel \mu,{k+\frac{1}{2},l-\frac{1}{2}}} \left.\frac{\partial f}{\partial \mu} \right|_{k+\frac{1}{2},l-\frac{1}{2}}  + \kappa_{v_\parallel \mu,{k+\frac{1}{2},l+\frac{1}{2}}} \left.\frac{\partial f}{\partial \mu} \right|_{k+\frac{1}{2},l+\frac{1}{2}}\right),\label{eqn:fp_diff_2}
\end{align}
where
\begin{align}
\left.\frac{\partial f}{\partial \mu} \right|_{k+\frac{1}{2},l\pm\frac{1}{2}} = \frac{1}{2} \left( \left.\frac{\partial f}{\partial \mu} \right|_{k,l\pm\frac{1}{2}} + \left.\frac{\partial f}{\partial \mu} \right|_{k+1,l\pm\frac{1}{2}} \right).\label{eqn:fp_diff_3}
\end{align}
The derivatives at the cell faces in (\ref{eqn:fp_diff_1}) and (\ref{eqn:fp_diff_3}) are computed using fourth-order central differences:
\begin{align}
\left.\frac{\partial f}{\partial v_\parallel}\right|_{k+\frac{1}{2},l} = \frac{1}{\Delta v_\parallel} \left( \frac{1}{24}f_{k-1,l} - \frac{9}{8} f_{k,l} + \frac{9}{8} f_{k+1,l} - \frac{1}{24} f_{k+2,l} \right),
\end{align}
and similarly for $\left.{\partial f}/{\partial \mu}\right|_{k,l+\frac{1}{2}}$. 
The advective and diffusive coefficients in (\ref{eqn:fp_adv_1}), (\ref{eqn:fp_diff_1}), and (\ref{eqn:fp_diff_2}) are related to the Rosenbluth potentials through (\ref{eqn:fpcoeffs}). Appendix~\ref{app:rosenbluth} describes the computation of the Rosenbluth potentials from the distribution function by discretizing and solving (\ref{eqn:rosenbluth}), which is defined on the infinite velocity domain but is solved on a truncated numerical domain. Our current implementation is 
second-order accurate, and therefore, the coefficients in (\ref{eqn:fp_adv_1}), (\ref{eqn:fp_diff_1}), and (\ref{eqn:fp_diff_2}) are calculated by discretizing (\ref{eqn:fpcoeffs}) with second-order finite differences:
\begin{subequations}
\begin{align}
\sigma_{v_\parallel,k+\frac{1}{2},l} =& \frac{1}{\Delta v_\parallel} \left( - \varphi_{k,l} + \varphi_{k+1,l} \right),\\
\kappa_{v_\parallel v_\parallel, k+\frac{1}{2},l} =& -\frac{1}{2}\left[ \frac{\varrho_{k-1,l} - 2 \varrho_{k,l} + \varrho_{k+1,l}}{\Delta v_\parallel^2} + \frac{\varrho_{k,l} - 2 \varrho_{k+1,l} + \varrho_{k+2,l}}{\Delta v_\parallel^2}\right],\\
\kappa_{v_\parallel\mu, k+\frac{1}{2},l+\frac{1}{2}} =& -\left(\frac{2\mu_{l+\frac{1}{2}}m}{B}\right) \frac{ \varrho_{k,l} - \varrho_{k+1,l} - \varrho_{k,l+1} + \varrho_{k+1,l+1} }{\Delta v_\parallel \Delta\mu}.
\label{eqn:fp_coeffs_3}
\end{align}
\end{subequations}
The magnetic field magnitude $B$ in (\ref{eqn:fp_coeffs_3}) varies in the configuration space only. Overall, the algorithm to compute $\hat{\mathcal{C}}\left(\bar{\bf f}\right)$ in (\ref{eqn:gk_semidisc}) is second-order accurate and $q=2$ in (\ref{eqn:semidisc_collision}).

\paragraph{Mass and Energy Conservation} The Fokker-Planck collision term conserves mass, momentum, and energy in its analytical form, and it is important that these quantities are discretely conserved as well. Mass conservation is expressed as
\begin{align}
\int\displaylimits_{\Omega_{\bf v}} c\left(f\right) d{\bf v} = 0 \Rightarrow \int\displaylimits_{\partial\Omega_{\bf v}} \overrightarrow{\Gamma} \cdot d{\bf s} = 0\label{eqn:fp_mass_cons}
\end{align}
where $\Omega_{\bf v} \equiv \left[ -v_{\parallel,\max}, v_{\parallel,\max} \right] \times \left[0, \mu_{\max} \right]$ is the velocity space domain, and ${\bf v} \equiv \left(v_\parallel, \mu\right)$ is the velocity space coordinate. 
Equation (\ref{eqn:fp_mass_cons}) is satisfied if the normal flux $\overrightarrow{\Gamma}\cdot d{\bf s}$ is zero over the entire velocity space boundary, and this is enforced by setting the coefficients $\sigma_{v_\parallel}$, $\sigma_{\mu}$, $\kappa_{v_\parallel v_\parallel}$, $\kappa_{\mu \mu}$, and $\kappa_{v_\parallel \mu}$ at the velocity boundaries to zero. Energy conservation is expressed as
\begin{align}
\mathcal{E}_{\left\{c\left(f\right)\right\}} \left({\bf R}\right) \equiv \int\displaylimits_{\Omega_{\bf v}} c\left(f\right)\left(v_\parallel^2 + \mu B\right) d{\bf v} = 0, \label{eqn:fp_energy_cons}
\end{align}
and is enforced using the approach described in~\cite{taitanoetal2015}, adapted for Cartesian velocity coordinates. Let
\begin{align}
c^{\left(a\right)} \left( f \right) = \nu_c \nabla_{\left(v_\parallel,\mu\right)} \cdot \left[ \overrightarrow{\sigma} f \right],\ \ \ 
c^{\left(d\right)} \left( f \right) = \nu_c \nabla_{\left(v_\parallel,\mu\right)} \cdot \left[ \overleftrightarrow{\boldsymbol\kappa} \nabla_{\left(v_\parallel,\mu\right)} f \right],
\end{align}
and let $\hat{\mathcal{C}}^{\left(a\right)} \left( \bar{\bf f} \right)$ and $\hat{\mathcal{C}}^{\left(d\right)} \left( \bar{\bf f} \right)$ be the corresponding spatially discretized terms. Equation (\ref{eqn:fp_energy_cons}) implies
\begin{align}
{\mathcal{E}_{\left\{ c^{\left(d\right)} \left( f \right) \right\}} \left({\bf R}\right)} = - { \mathcal{E}_{\left\{ c^{\left(a\right)} \left( f \right) \right\}} \left({\bf R}\right) }.
\end{align}
We define 
\begin{align}
\nu_{\mathcal{E}} \left({\bf R}\right) = - {\mathcal{E}_{\left\{ \hat{\mathcal{C}}^{\left(a\right)} \left( \bar{\bf f} \right) \right\}} \left({\bf R}\right)} / { \mathcal{E}_{\left\{ \hat{\mathcal{C}}^{\left(d\right)} \left( \bar{\bf f} \right) \right\}} \left({\bf R}\right) },
\end{align}
and modify (\ref{eqn:fp}) as
\begin{align}
c \left( f \right) = \nu_c \nabla_{\left(v_\parallel,\mu\right)} \cdot \overrightarrow{\Gamma}, \ \ \ \ \ \overrightarrow{\Gamma} &= \overrightarrow{\sigma} f + \nu_{\mathcal{E}} \left({\bf R}\right) \overleftrightarrow{\boldsymbol\kappa} \nabla_{\left(v_\parallel,\mu\right)} f
\label{eqn:modified_fp}
\end{align}
We note that $\nu_{\mathcal{E}}\left({\bf R}\right) \rightarrow -1$ as $\Delta v_\parallel, \Delta\mu \rightarrow 0$ for a consistent discretization of the collision term, and this factor counteracts the energy imbalance in the collision term resulting from the discretization errors. Equation~(\ref{eqn:modified_fp}) is discretized as decribed in the preceding discussions. Discrete enforcement of momentum conservation is also discussed in~\cite{taitanoetal2015}; however, the numerical test cases considered in this paper have negligible average velocities and yielded stable and convergent results without momentum conservation enforcement. In our implementation, discrete energy conservation is necessary to preserve a Maxwellian solution; without the modification described by (\ref{eqn:modified_fp}), unphysical cooling of the Maxwellian solution is observed that is alleviated by refining the grid in the velocity space. This behavior is similar to that reported in~\cite{taitanoetal2015}.

\section{Time Integration}
\label{sec:time}

Equation~(\ref{eqn:gk_semidisc}) represents a system of ODEs in time that are solved using high-order, multistage explicit and IMEX time integrators in COGENT. The plasma density and temperature vary significantly in the edge region along the radial direction, and as a result, the stiffness of (\ref{eqn:gk_semidisc}) changes substantially from the inner edge region (adjacent to the core) to the outer edge. We use a simple problem to analyze the eigenvalues of the RHS of (\ref{eqn:gk_semidisc}). The two-dimensional (in space) Cartesian domain is ${\bf R} \equiv \left(r,\theta\right) \equiv \left(x,y\right)$ with specified density and temperature and a Maxwellian distribution function. The periodic domain is $0 \leq x,y < 1$, where $x = \tilde{x}/L_{\rm ref}, y = \tilde{y}/L_{\rm ref}$ are the non-dimensional spatial coordinates; $\tilde{x},\tilde{y}$ are the physical coordinates; and $L_{\rm ref} = 1\,{\rm m}$ is the reference length. A uniform density $n = 1$ is specified, and the temperature is specified as
\begin{align}
T = 1 + 0.1\cos{\left(2\pi y\right)},
\end{align}
where $n$ and $T$ are nondimensional quantities. The magnetic field is $\left(B_x,B_y,B_z\right) = \left(0,0.2,2\right)$ where $z$ is normal to the plane of the domain, and the reference magnetic field is $B_{\rm ref} = 1\,{\rm T}$. The reference mass is specified as the proton mass, i.e., $m_{\rm ref} = 1.6726 \times 10^{-24}\,{\rm g}$. We consider two values for the reference density and temperature, corresponding to two extremes of the edge region: the hot near-core region with $n_{\rm ref} = 10^{20}\,{\rm m}^{-3},\ T_{\rm ref} = 500\,{\rm eV}$, and the cold edge region with $n_{\rm ref} = 10^{19}\,{\rm m}^{-3},\ T_{\rm ref} = 20\,{\rm eV}$. These values are based on measurements in the edge region of a DIII-D tokamak~\cite{porteretal2000}. 

\begin{figure}[t]
\subfigure[Hot near-core]{\includegraphics[width=0.49\textwidth]{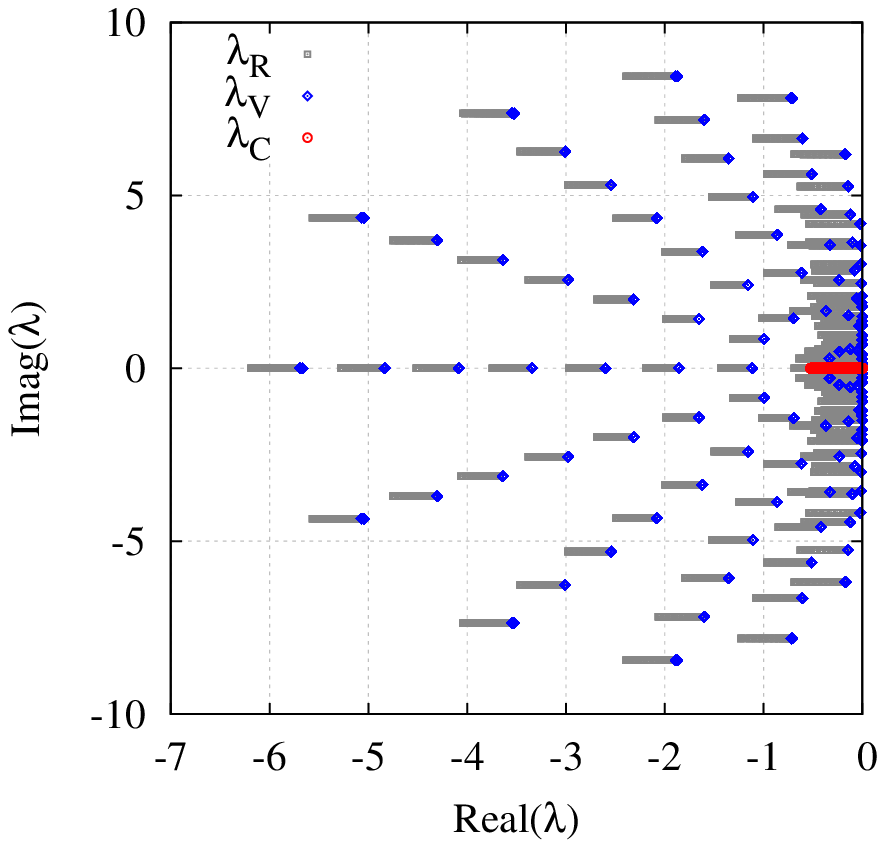}\label{fig:eigenvalues_hotcore}}
\subfigure[Cold edge]{\includegraphics[width=0.49\textwidth]{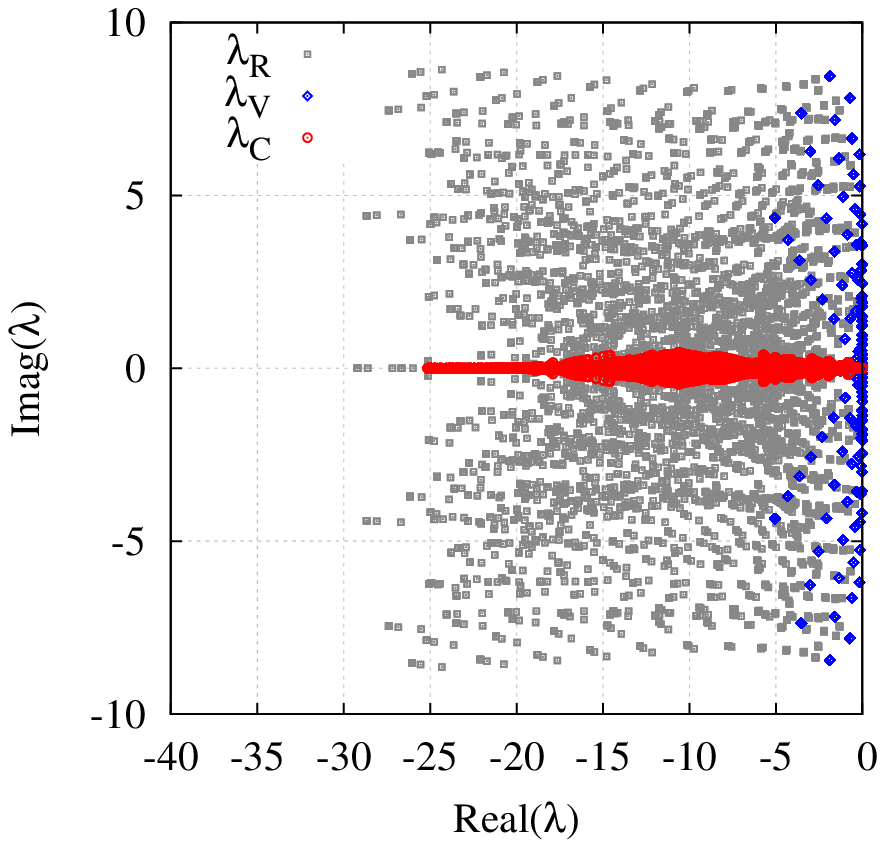}\label{fig:eigenvalues_coldedge}}
\caption{Eigenvalues of the RHS of (\ref{eqn:gk_semidisc}) ($\lambda_{\rm R}$) and its constituent terms -- the Vlasov ($\lambda_{\rm V}$) and collision terms ($\lambda_{\rm C}$). (a) In the hot near-core region, the collisional eigenvalues (red) are much smaller than the Vlasov eigenvalues (blue). (b) However, in the cold edge region, the former are much larger in magnitude than the latter. Note the change in the horizontal scale.}
\label{fig:eigenvalues}
\end{figure}

The Jacobians of the RHS of (\ref{eqn:gk_semidisc}) and its constituents $\hat{\mathcal{V}}$ and $\hat{\mathcal{C}}$ are computed by applying finite differences on these operators, which scales as $\mathcal{O}\left(N^2\right)$ operations where $N$ is the total number of grid points. Therefore, a very coarse grid with $\left(N_x, N_y, N_{v_\parallel}, N_\mu\right) = \left(6,16,16,12\right)$ points is used to discretize the domain for this analysis.
Figure~\ref{fig:eigenvalues} shows the eigenvalues of the RHS of (\ref{eqn:gk_semidisc}) as well as its constituents evaluated separately:
\begin{align}
\lambda_{\rm R} = \lambda\left( \left[\frac{d \hat{\mathcal{R}}\left(\bar{\bf f}\right)}{d \bar{\bf f}}\right] \right),\ 
\lambda_{\rm V} = \lambda\left( \left[\frac{d \hat{\mathcal{V}}\left(\bar{\bf f}\right)}{d \bar{\bf f}}\right] \right),\ 
\lambda_{\rm C} = \lambda\left( \left[\frac{d \hat{\mathcal{C}}\left(\bar{\bf f}\right)}{d \bar{\bf f}}\right] \right),
\end{align}
where $\hat{\mathcal{R}}\left(\bar{\bf f}\right) = \hat{\mathcal{V}}\left(\bar{\bf f}\right) + \hat{\mathcal{C}}\left(\bar{\bf f}\right)$ and $\lambda\left(\left[\cdot\right]\right)$ are the eigenvalues of $\left[\cdot\right]$. In the hot near-core region, shown in Figure~\ref{fig:eigenvalues_hotcore}, the eigenvalues of the Vlasov term dominate the spectrum, and thus, the overall system of equations is not stiff with respect to the Vlasov time scale. However, in the cold edge region, shown in Figure~\ref{fig:eigenvalues_coldedge}, the strong collisionality results in the eigenvalues of $\hat{\mathcal{C}}$ being much larger than those of the eigenvalues of $\hat{\mathcal{V}}$. It should be noted that the magnitudes of the eigenvalues of $\hat{\mathcal{V}}$ are similar in the two figures.

The stable time step for an explicit time integration method is constrained by
\begin{align}
\lambda_{\rm R}\Delta t \in \left\{z : \mathcal{P}_{\rm exp} \left(z\right) \le 1 \right\},
\end{align}
where $\mathcal{P}$ represents the stability polynomial. It is evident from Figure~\ref{fig:eigenvalues} that, though an explicit method will allow time steps comparable to the Vlasov time scales in the near-core region, it will be constrained by the collisional time scales in the cold edge region. This fact motivates the use of the IMEX approach, where the collisions are integrated in time implicitly, while the Vlasov term is integrated explicitly; therefore, the stable time step for an IMEX method is constrained by
\begin{align}
\lambda_{\rm V}\Delta t \in \left\{z : \mathcal{P}_{\rm IMEX} \left(z\right) \le 1 \right\},
\end{align}
where $\mathcal{P}_{\rm IMEX}$ is the stability polynomial for the explicit component of the IMEX method. The IMEX approach allows time steps comparable to the Vlasov time scales in the entire edge region.

In this paper, we investigate high-order multistage ARK methods~\cite{kennedycarpenter} that are expressed in the Butcher tableaux~\cite{butcher2003} form as:
\begin{align}
\left(
\begin{array}{c|c}
  c_i & a_{ij}\\
\hline
  & b_j
\end{array},\ 
\begin{array}{c|c}
  \tilde{c}_i & \tilde{a}_{ij}\\
\hline
  & \tilde{b}_j
\end{array}; i,j = 1,\cdots,n_s
\right),\label{eqn:ark}
c_i = \sum_{j=1}^{n_s} a_{ij}, \tilde{c}_i = \sum_{j=1}^{n_s} \tilde{a}_{ij},
\end{align}
where $a_{ij},b_j,c_i$ define the explicit component of the IMEX method, with $a_{ij} = 0, j \ge i$; $\tilde{a}_{ij},\tilde{b}_j,\tilde{c}_i$ define the implicit component, with $\tilde{a}_{ij} = 0, j > i$; and $n_s$ is the number of stages. The implicit components of the methods considered in this paper are $L$-stable, explicit-first-stage, single-diagonal-coefficient implicit Runge-Kutta (ESDIRK) methods, and the coefficients satisfy $\tilde{a}_{11} = 0;\ \tilde{a}_{ii} = \gamma, i=2,\cdots,n_s$, where $\gamma$ is a constant. 
In addition, the methods are conservative, and $b_i = \tilde{b}_i, \forall i$.
A time step of the ARK methods applied to Eq.~(\ref{eqn:gk_semidisc}) is expressed as follows:
\begin{subequations}
\label{eqn:ARK}
\begin{align}
\bar{\bf f}^{\left(i\right)} &= \bar{\bf f}_n + \Delta t \left\{ \sum_{j=1}^{i-1} a_{ij} \hat{\mathcal{V}}\left(\bar{\bf f}^{\left(j\right)}\right) + \sum_{j=1}^{i} \tilde{a}_{ij} \hat{\mathcal{C}}\left(\bar{\bf f}^{\left(j\right)}\right) \right\},\ \ \ i=1,\cdots,n_s,\label{eqn:ark_stage}\\
\bar{\bf f}_{n+1} &= \bar{\bf f}_n + \Delta t \sum_{i=1}^{n_s} \left\{ b_i \hat{\mathcal{V}}\left(\bar{\bf f}^{\left(i\right)}\right) + \tilde{b}_{i} \hat{\mathcal{C}}\left(\bar{\bf f}^{\left(i\right)}\right) \right\}\label{eqn:ark_step},
\end{align}
\end{subequations}
where $\Delta t$ is the time step, $n$ denotes the time level ($\bar{\bf f}_n = \bar{\bf f}\left(t_n\right), t_n = t_0 + n\Delta t$), and $\bar{\bf f}^{\left(i,j\right)}$ are the stage solutions. The second--order (three--stage) (``ARK2")~\cite{giraldokellyconsta2013}, the third-order (four-stage) (``ARK3")~\cite{kennedycarpenter}, and the fourth-order (six-stage) (``ARK4")~\cite{kennedycarpenter} are implemented.

Equation~(\ref{eqn:ark_stage}) is a nonlinear system of equations for $i>1$ that is expressed as
\begin{align}
\mathcal{F}\left({\bf y}\right) \equiv \alpha {\bf y} - \hat{\mathcal{C}}\left({\bf y}\right) - {\bf r} = 0, \label{eqn:nonlinear}
\end{align}
where ${\bf y} \equiv  \bar{\bf f}^{\left(i\right)}$ is the unknown stage solution and
\begin{align}
\alpha = \frac{1}{\gamma\Delta t},\ \ \ 
{\bf r} = \frac{1}{\gamma\Delta t} \left[ \bar{\bf f}_n + \Delta t \sum_{j=1}^{i-1} \left\{ a_{ij} \hat{\mathcal{V}}\left(\bar{\bf f}^{\left(j\right)}\right) + \tilde{a}_{ij} \hat{\mathcal{C}}\left(\bar{\bf f}^{\left(j\right)}\right) \right\} \right].
\end{align}
We use the Jacobian--Free Newton-Krylov (JFNK) approach~\cite{knollkeyes2004} to solve Eq.~(\ref{eqn:nonlinear}). The inexact Newton's method~\cite{dennisschnabel} is used to solve the nonlinear equation, where the initial guess is the solution of the previous stage ${\bf y}_0 \equiv \bar{\bf f}^{\left(i\right)}_0 = \bar{\bf f}^{\left(i-1\right)}$ (note that the first stage is always explicit), and the $k$-th Newton step is
\begin{align}
{\bf y}_{k+1} = {\bf y}_k - \mathcal{J}\left({\bf y}_k\right)^{-1}\mathcal{F}\left({\bf y}_k\right);\ \ \ \mathcal{J}\left({\bf y}\right) = \frac {d \mathcal{F}\left({\bf y}\right)} {d {\bf y}} = \alpha \mathbb{I} - \frac {d \hat{\mathcal{C}}\left({\bf y}\right)} {d {\bf y}},\label{eqn:newton_step}
\end{align}
where $\mathbb{I}$ is the identity matrix and $\mathcal{J}$ is the Jacobian of $\mathcal{F}$.
The exit criterion is
\begin{align}
\|\mathcal{F}\left({\bf y}_k\right)\|_2 \le \max\left({\epsilon}_r\|\mathcal{F}\left({\bf y}_0\right)\|_2, {\epsilon}_a\right)\ {\rm or}\ \|\mathcal{J}^{-1}\mathcal{F}\left({\bf y}_k\right)\|_2 \le {\epsilon}_s,
\end{align}
where ${\epsilon}_a$ and ${\epsilon}_r$ are the absolute and relative tolerances for the Newton solver, respectively, and ${\epsilon}_s$ is the step size tolerance.
Equation~(\ref{eqn:newton_step}) requires the solution to the linear system of equations,
\begin{align}\label{eqn:linear_system}
\left[ \mathcal{J}\left({\bf y}_k\right) \right] {\bf x} = \mathcal{F}\left({\bf y}_k\right),
\end{align}
which is solved using the preconditioned generalized minimum residual (GMRES) method~\cite{saad,saad1986}. The GMRES solver only requires the action of the matrix $\mathcal{J}$ on a vector, which is approximated by computing the directional derivative~\cite{pernicewalker},
\begin{align}
\mathcal{J}\left({\bf y}_k\right) {\bf x} \approx \frac{1}{\epsilon} \left[ \mathcal{F}\left({\bf y}_k + \epsilon{\bf x}\right) - \mathcal{F}\left({\bf y}_k\right) \right] = \alpha{\bf x} - \frac{1}{\epsilon} \left[ \hat{\mathcal{C}}\left({\bf y}_k + \epsilon{\bf x}\right) - \hat{\mathcal{C}}\left({\bf y}_k\right) \right],
\end{align}
where $\epsilon = {\sqrt{\epsilon_m\left(1+\|{\bf y}_k\|_2\right)}}/{\|{\bf x}\|_2}$ and $\epsilon_m$ is machine round-off error (taken as $10^{-14}$ in our algorithm). The exit criterion for the GMRES solver is
\begin{align}
\|{\bf r}_l\|_2 \le \max\left( \tilde{\epsilon}_a, \tilde{\epsilon}_r\|{\bf r}_0\|_2 \right),\ \ 
{\bf r}_l = \mathcal{J}\left({\bf y}_k\right) {\bf x}_l + \mathcal{F}\left({\bf y}_k\right),
\end{align}
where ${\bf x}_l$ is the solution at the $l$-th GMRES iteration, and ${\bf r}_0 = \mathcal{F}\left({\bf y}_k\right)$ since ${\bf x}_0 = {\bf 0}$.

The preconditioned GMRES algorithm uses a preconditioning matrix to accelerate convergence. Equation~(\ref{eqn:linear_system}) is rewritten as
\begin{align}\label{eqn:linear_system_precon}
\left[ \mathcal{J}\left({\bf y}_k\right) \right] \left[ \mathcal{P}\left({\bf y}_k\right) \right] \left[ \mathcal{P}\left({\bf y}_k\right) \right]^{-1} {\bf x} = \mathcal{F}\left({\bf y}_k\right),
\end{align}
where $\mathcal{P} \approx \mathcal{J}$ is the preconditioning matrix. Although the Jacobian $\mathcal{J}$ is not assembled, the preconditioning matrix $\mathcal{P}$ is assembled and stored as a banded matrix. The preconditioner is constructed as
\begin{align}
\mathcal{P}\left({\bf y}\right) = \alpha \mathcal{I} - \frac {d \bar{\mathcal{C}}\left({\bf y}\right)} {d {\bf y}},\ \ 
\bar{\mathcal{C}}\left(\bar{\bf f}\right) = \mathcal{C}\left(\bar{\bf f}\right) + \mathcal{O}\left( \Delta v_\parallel^r, \Delta \mu^r \right),\ r < q.
\end{align}
Thus, the preconditioning matrix is defined as the Jacobian of a lower--order discretization of the collision term. In our current implementation, $\bar{\mathcal{C}}$ is constructed by reconstructing $f_{k+\frac{1}{2},l}$ in (\ref{eqn:fp_adv_1}) with the first-order upwind method and discretizing $\left.{\partial_{v_\parallel} f}\right|_{k+\frac{1}{2},l}$ and $\left.{\partial_{\mu} f}\right|_{k,l+\frac{1}{2}}$ in (\ref{eqn:fp_diff_1}) and (\ref{eqn:fp_diff_3}) with second-order central differences. In addition, the conversions from cell-centered to cell-averaged solutions and vice-versa using (\ref{eqn:conversions}) are neglected, which results in a nine-banded sparse matrix. The Gauss--Seidel method is used to invert it. Though this preconditioning approach is satisfactory for the cases presented in this paper, we are exploring more efficient approaches presented in the literature~\cite{taitanoetal2015}.

\section{Results}
\label{sec:results}

In this section, we test the accuracy, convergence, and computational cost of the IMEX time integration methods by applying them to two cases that are representative of the tokamak edge region. The edge is characterized by a steep decrease in the plasma density and temperature in the outward radial direction~\cite{porteretal2000} that results in an increase in the collisionality from the near-core region to the edge. Our current implementation is insufficient to resolve significant variations in temperature due to the use of a fixed velocity grid in the entire configuration space domain; implementation of adaptive velocity grids based on the local thermal velocity will be pursued in future studies. We investigate our time integration methods for constant and varying collisionality by considering two Cartesian, two-dimensional heat transport problems where ${\bf R} \equiv \left(x,y\right)$. Both of these problems involve the thermal equilibration of an initially sinusoidal temperature profile while a specified electrostatic potential drives the flow. The first problem is essentially one-dimensional with zero gradients along the $x$ dimension, and the collisionality does not vary significantly through the domain. The second problem introduces significantly varying collisionality along the $x$ dimension (similar to the variation at the tokamak edge) by specifying an appropriate density profile. The performance of the IMEX methods is analyzed for both these problems and compared with that of the explicit RK methods.

The cases and the results in this section are described in terms of the nondimensional variables described in Section~\ref{sec:govern}, unless noted otherwise. The reference quantities (dimensional) and their values are
\begin{align}\label{eqn:ref}
\begin{array}{lll}
L_{\rm ref} = 1\,{\rm m}, & n_{\rm ref} = 10^{20}\,{\rm m}^{-3}, & T_{\rm ref} = 20\,{\rm eV}, \\
m_{\rm ref} = 1.6726\times 10^{-24}\,{\rm g}, & B_{\rm ref} = 1\,{\rm T}, & v_{\rm ref} = 4.377 \times 10^6\,{\rm cm}/{\rm s}, \\
t_{\rm ref} = 2.3\times 10^{-5}\,{\rm s}, & \mu_{\rm ref} = 1.602\times 10^{-15}\,{\rm g}\,{\rm cm}^{2}\,{\rm s}^{-2}\,{\rm G}^{-1} & \phi_{\rm ref} = 20\,{\rm V},
\end{array}
\end{align}
and the nondimensional values in the subsequent discussions should be multiplied by the corresponding reference values in (\ref{eqn:ref}) to calculate the dimensional values.

COGENT is implemented for distributed-memory parallel computations using the MPI library, and the simulations discussed in this paper are run on a Linux cluster where each node has two Intel Xeon $8$-core processors with $2.6\,{\rm GHz}$ clock speed and $32\,{\rm GB}$ of memory.

\subsection{Case 1: Uniform Collisionality}
\label{sec:case1}

\begin{figure}[t]
\subfigure[Density $n(y)$]{\includegraphics[width=0.49\textwidth]{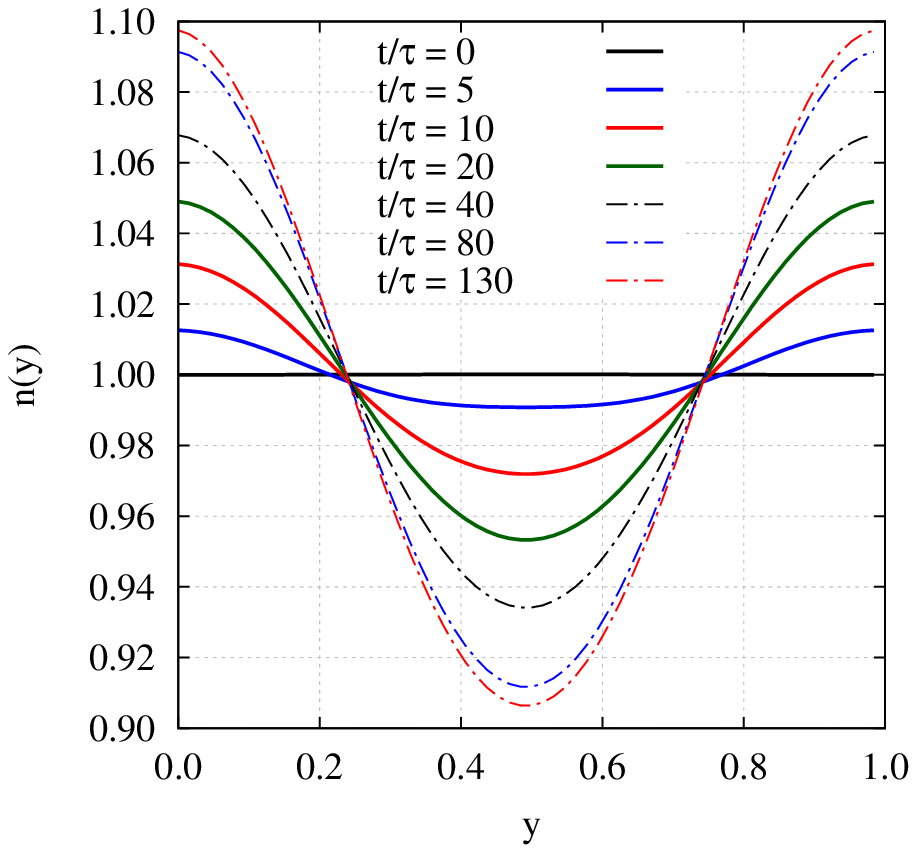}\label{fig:case1_density}}
\subfigure[Temperature $T(y)$]{\includegraphics[width=0.49\textwidth]{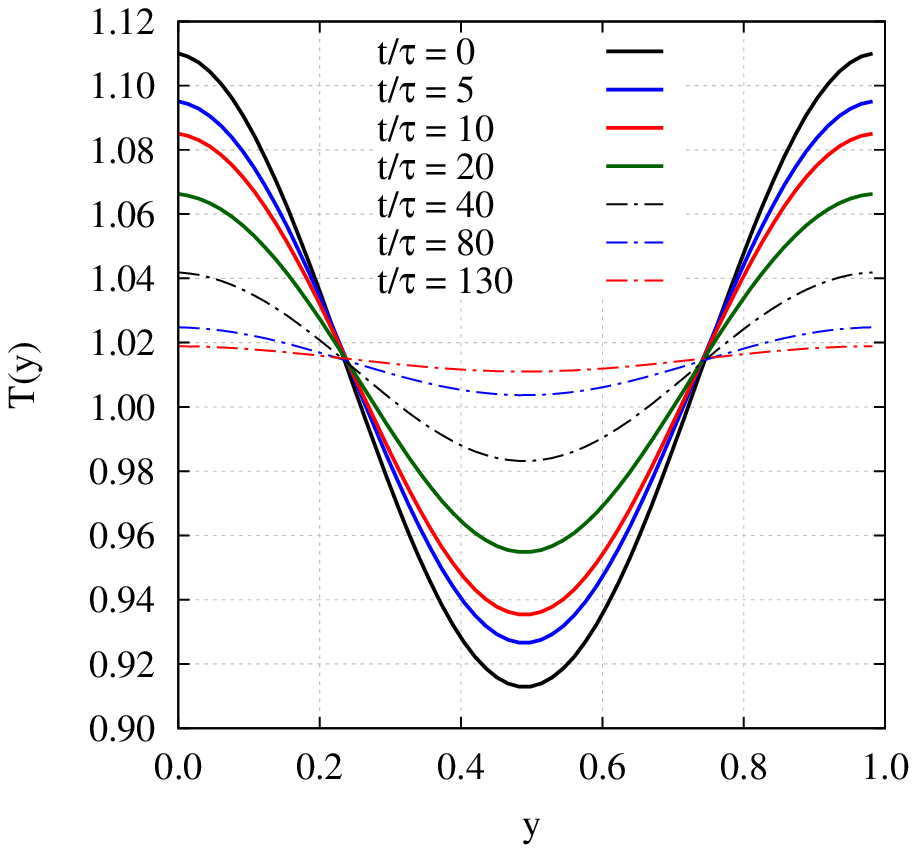}\label{fig:case1_temperature}}
\caption{Evolution of density and temperature in time. The constant initial density assumes a cosine-like profile to balance the specified electrostatic potential. Thermal equilibration drives the temperature to a constant value.}
\label{fig:case1_physics}
\begin{center}
\includegraphics[width=0.9\textwidth]{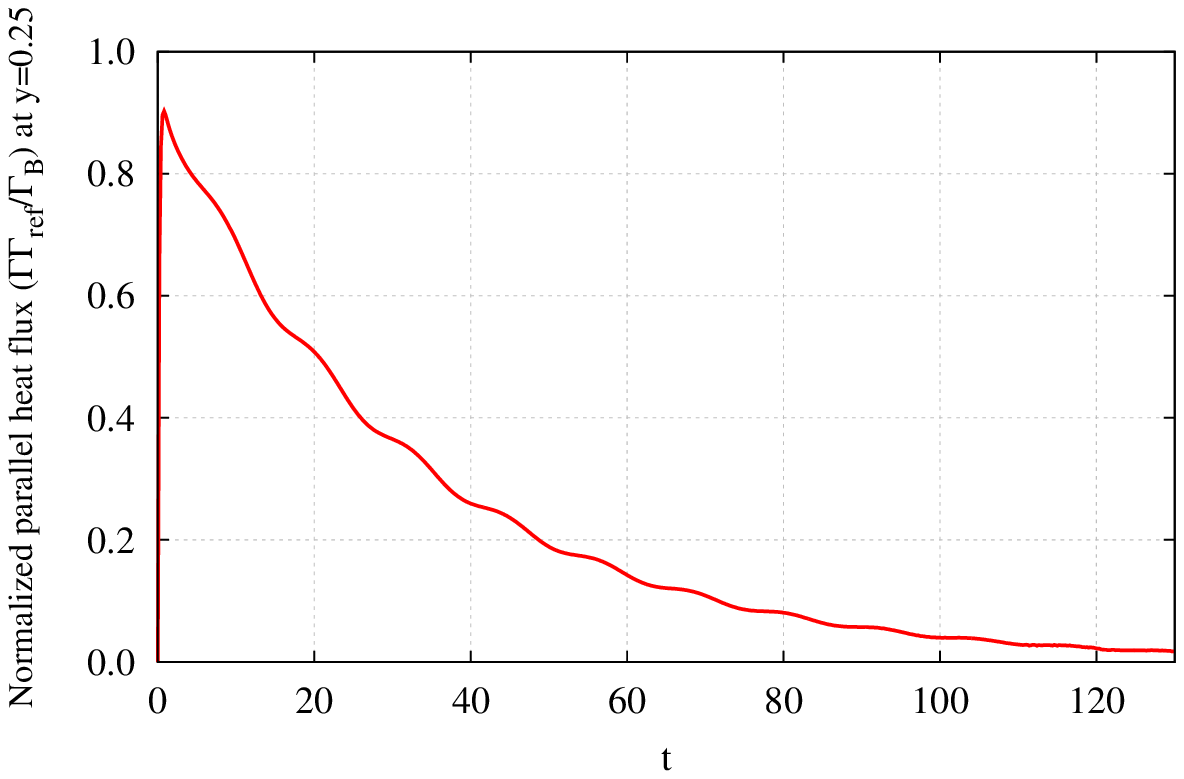}
\end{center}
\caption{Parallel heat flux as a function of time at $y=0.25$: The heat flux attains its ``quasi-steady" value consistent with the temperature gradient at an instant in time. As the temperature equilibriates, it decays to zero.}
\label{fig:case1_phf}
\end{figure}

This case simulates the parallel heat transport over a two-dimensional (in configuration space) slab where the collisionality does not vary significantly throughout the domain. The configuration space domain is $0 \le x,y \le L$, where $L=1$, and the velocity space domain is $-3.5 \le v_\parallel \le 3.5,\ 0 \le \mu \le 10$. The initial solution is the Maxwellian distribution in the velocity space with uniform density $n\left(x,y,t=0\right) = 1$, and the initial temperature profile is specified as
\begin{align}
T(x,y,t=0) \equiv T_0(y) = 1 + 0.1 \cos{\left(2\pi y\right)}.
\end{align}
A steady electrostatic potential $\phi\left(x,y,t\right) = -0.1\cos{\left(2\pi y\right)}$ and a constant magnetic field $B_y = 0.2, B_z = 2$ are applied throughout the simulation. Periodic boundary conditions are applied at all boundaries. Flow gradients along $x$ are zero at all times in the simulation, and this case is essentially one-dimensional. The species charge and mass correspond to ionized hydrogen. The collision time is defined as
\begin{align}
\tilde{\tau} = \frac{3}{4} \sqrt{\frac{\tilde{m}\tilde{T}^3}{\pi}} \frac{1}{\tilde{n}e^4 \lambda_c} = 2.4\times 10^{-6}\,{\rm s} \Rightarrow \tau = \frac{\tilde{\tau}}{t_{\rm ref}} = 0.104,
\end{align}
where $\tilde{\left(\cdot\right)}$ are dimensional quantities and $\tau$ is the nondimensional collision time. The Coulomb logarithm is specified as $\lambda_c = 11$~\cite{nrlformulary}. The collisional mean free path is $\lambda = v_{\rm th} \tau$, where $v_{\rm th} = \sqrt{2{T}/{m}}$ is the thermal velocity, and the characteristic length scale is $k_\parallel^{-1} = \left(B/B_y\right)\left(L/2\pi\right)$; consequently, the ratio of the particle mean free path to the characteristic length scale is
\begin{align}
k_\parallel \lambda = 0.065,
\end{align}
and therefore, the plasma is highly collisional for this case.

The solutions discussed in this section are obtained on a grid with $\left(N_x, N_y, N_{v_\parallel}, N_\mu\right) = \left(6, 64, 36, 24\right)$ points. Figure~\ref{fig:case1_physics} shows the evolution of the density and temperature in time. This solution is obtained with the ARK4 method with a time step of $\Delta t = 0.05$ that results in a Vlasov CFL of $\sigma_{\rm V} \sim 1.1$ and a collision CFL of $\sigma_{\rm C} \sim 12.9$, defined as follows:
\begin{subequations}\label{eqn:cfl}
\begin{align}
\sigma_{\rm V} &= \max \left\{ \frac{\left(\dot{\bf R}\cdot\hat{\bf x}\right)\Delta t}{\Delta x}, \frac{\left(\dot{\bf R}\cdot \hat{\bf y}\right) \Delta t}{\Delta y}, \frac{\dot{v}_\parallel \Delta t}{\Delta v_\parallel} \right\}, \\
\sigma_{\rm C} &= \max \left\{ \frac{\sigma_{v_\parallel}\Delta t}{\Delta v_\parallel}, \frac{\sigma_\mu\Delta t}{\Delta \mu}, \frac{\kappa_{v_\parallel v_\parallel}\Delta t}{\Delta v_\parallel^2}, \frac{\kappa_{\mu\mu}\Delta t}{\Delta \mu^2}, \frac{\kappa_{v_\parallel\mu}\Delta t}{\Delta v_\parallel \Delta \mu} \right\}.
\end{align}
\end{subequations}
The simulation is run until a final time of $t_f=130$ on $192$ cores ($12$ nodes) with $8$ MPI ranks along $y$, $6$ ranks along $v_\parallel$, and $4$ ranks along $\mu$. The specified electrostatic potential forces the initially constant density to assume a sinusoidal profile while thermal equilibration drives the temperature to a constant value. These processes occur on the slow transport time scale ($\tau_T \approx mL^2/2\tau T$). Figure~\ref{fig:case1_phf} shows the normalized parallel heat flux ($\Gamma\Gamma_{\rm ref}/\Gamma_{\rm B}$) at $y=0.25$ as a function of time, where $\Gamma_{\rm B} = 1.87\times 10^8\,{\rm g}\,{\rm s}^{-3}$ is the heat flux computed by the Braginskii model~\cite{braginskii},
\begin{align}
\Gamma\left({\bf R}\right) = \int\int \left\{ v_\parallel - \overline{v_\parallel}\left({\bf R}\right) \right\} \frac{\left\{v_\parallel - \overline{v_\parallel}\left({\bf R}\right)\right\}^2 + \mu B} {2} f\left({\bf R},v_\parallel,\mu\right) B_\parallel^*\left({\bf R}, v_\parallel \right)dv_\parallel d \mu
\end{align}
is the nondimensional parallel heat flux, and 
\begin{equation}\Gamma_{\rm ref} = n_{\rm ref} T_{\rm ref}^{\frac{3}{2}} m_{\rm ref}^{-\frac{1}{2}} = 1.4\times 10^{10}\,{\rm g}\,{\rm s}^{-3}
\end{equation}
is the reference heat flux.
At a given time, the collisions drive the heat flux to a value consistent with the local temperature gradient at the collisional time scale $\tau$. As the temperature assumes a constant value, the heat flux decays to zero, which happens on the transport time scale. Thus, this problem demonstrates the disparate time scales resulting from a highly-collisional plasma.

\begin{figure}[t]
\subfigure[Error vs. CFL]                     {\includegraphics[width=0.49\textwidth]{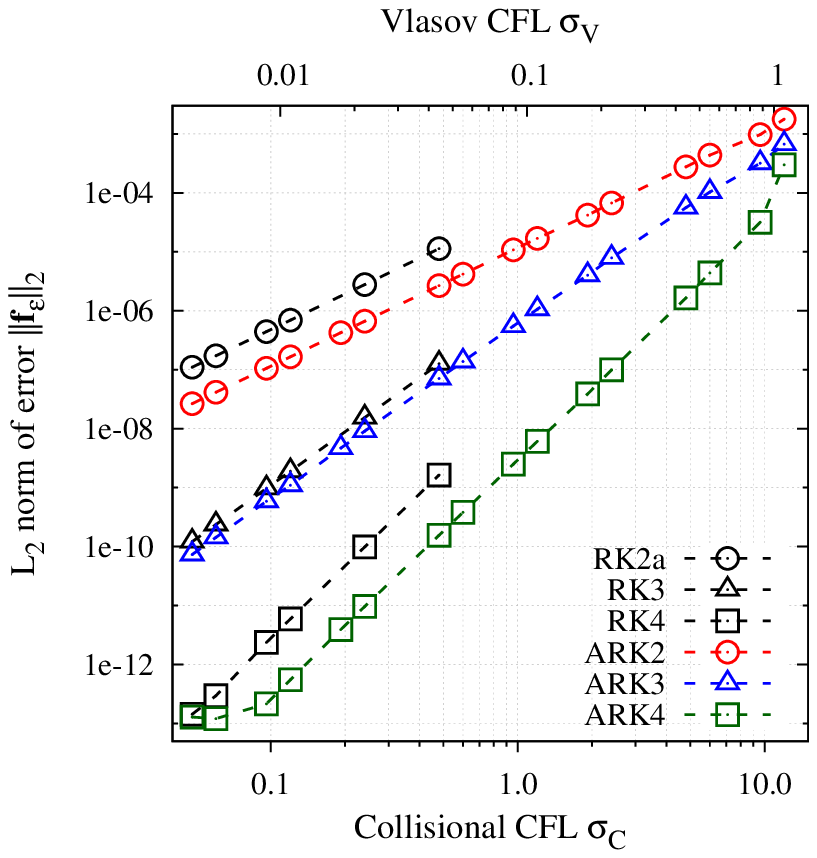}\label{fig:case1_errdt}}
\subfigure[Error vs. solver tolerance (ARK4) ]{\includegraphics[width=0.49\textwidth]{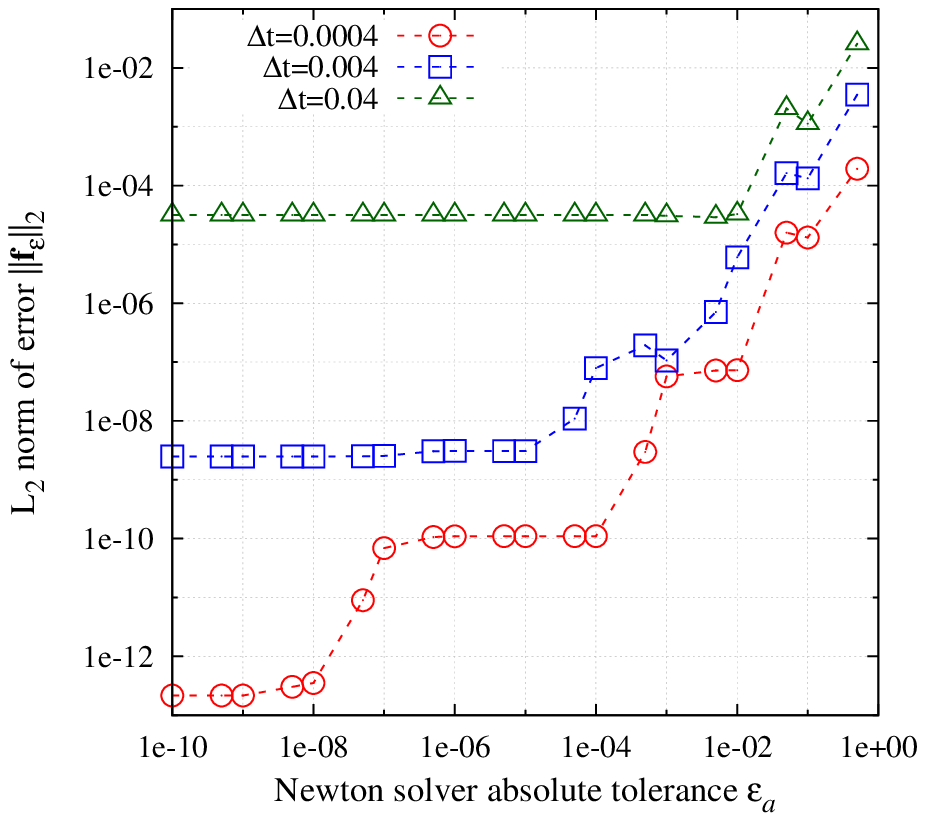}\label{fig:case1_errtol}}
\caption{ Accuracy and convergence: (a) The $L_2$ norm of the error as a function of the collisional and Vlasov CFL numbers. The time integration methods converge at their theoretical orders of convergence. (b) The error as a function of the Newton solver absolute tolerance for ARK4 at three values of $\Delta t$. A higher $\Delta t$ allows a more relaxed tolerance without increasing the truncation error.}
\label{fig:case1_convergence}
\subfigure[Error vs. number of function calls]{\includegraphics[width=0.49\textwidth]{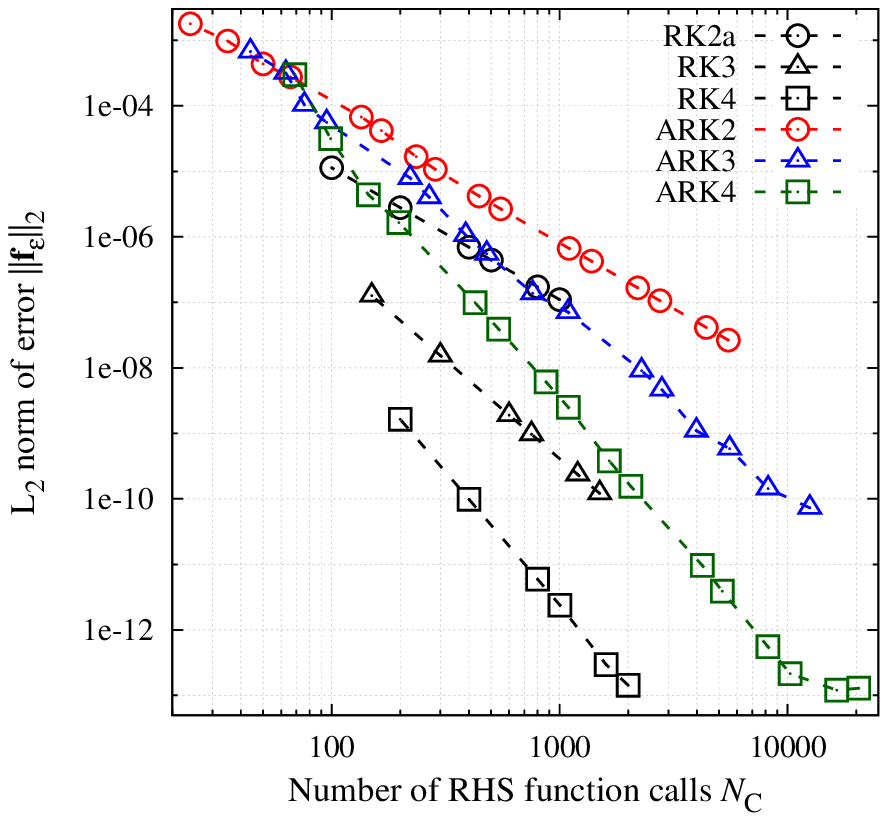}\label{fig:case1_errfunccount}}
\subfigure[Error vs. wall time]{\includegraphics[width=0.49\textwidth]{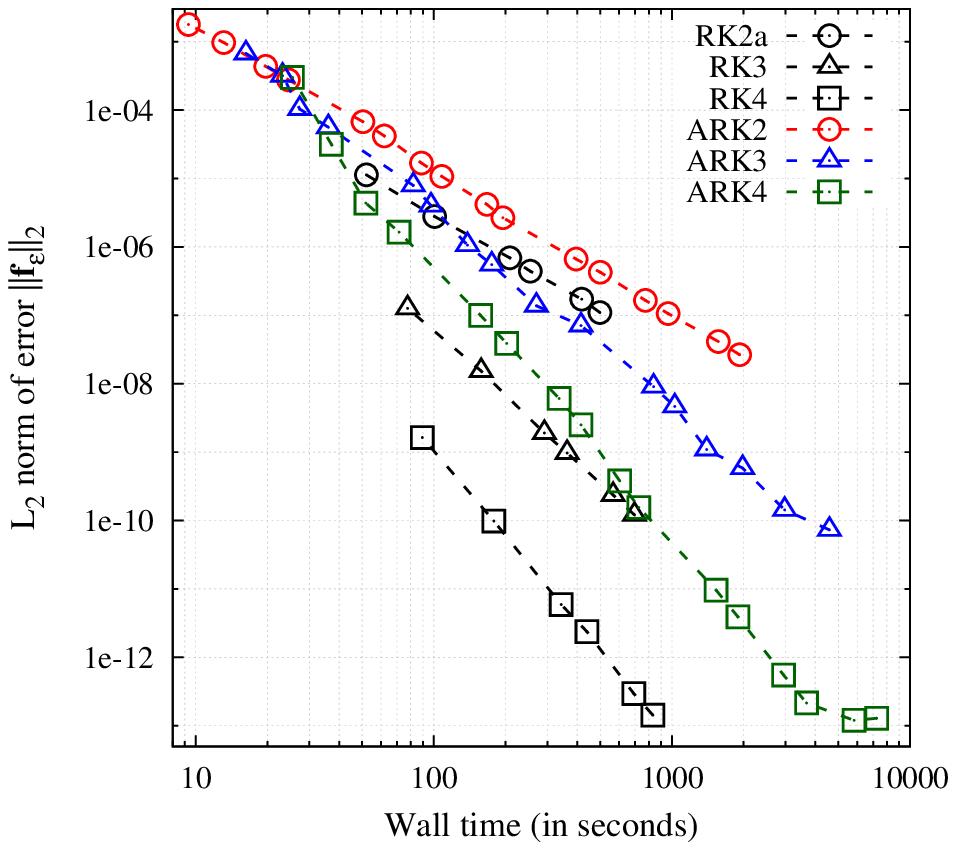}\label{fig:case1_errwalltime}}
\caption{Computational cost: (a) The error as a function of the number of calls to the collision operator $\hat{\mathcal{C}}$ and (b) the error as a function of the wall time in seconds. The IMEX approach allows significantly faster stable solutions compared with explicit time integration.}
\label{fig:case1_cost}
\end{figure}

Figure~\ref{fig:case1_errdt} shows the $L_2$ norm of the time integration error as a function of the collisional and Vlasov CFL numbers defined by (\ref{eqn:cfl}) for solutions obtained at a final time of $t_f=0.1$. The errors for the ARK methods are compared with those for explicit second-order (RK2a), third-order (RK3), and fourth-order (RK4) RK methods. The error is defined as
\begin{align}
{\bf f}_\epsilon = \bar{\bf f} - \bar{\bf f}_{{\rm RK4},\Delta t=10^{-5}},\label{eqn:errorvec}
\end{align}
where $\bar{\bf f}$ is the numerical solution and $\bar{\bf f}_{{\rm RK4},\Delta t=10^{-5}}$ is the reference solution obtained with the RK4 method with a time step of $\Delta t = 10^{-5}$ (corresponding to $\sigma_{\rm V} \sim 2.2 \times 10^{-4},\ \sigma_{\rm C} \sim 2.4 \times 10^{-3}$). The relative tolerance for the Newton solver is $\epsilon_r = 10^{-10}$ for all the solutions, and the absolute tolerance $\epsilon_a$ is specified as the maximum value $\epsilon_a^{\max}$ for which the error in solving the nonlinear system is less than the truncation error of the time integration method for a given $\Delta t$. Thus, it is a function of the method and the time step size and is defined as
\begin{align}
\epsilon_a^{\max} = \max{\left(\epsilon_\alpha\right)}\ \ \ {\rm such\ that}\ \ \ \| {\bf f}_{\epsilon, \epsilon_a} \|_2 = \| {\bf f}_{\epsilon, 10^{-10}} \|_2,\label{eqn:newton_tol_max}
\end{align}
where ${\bf f}_{\epsilon, \epsilon_a}$ is the error defined by (\ref{eqn:errorvec}) for a solution obtained with an absolute tolerance of $\epsilon_a$, and ${\bf f}_{\epsilon, 10^{-10}}$ is the error for a solution obtained with an absolute tolerance of $10^{-10}$. As an example, Figure~\ref{fig:case1_errtol} shows the $L_2$ norm of the error as a function of the absolute tolerance $\epsilon_a$ for the ARK4 method at three values of $\Delta t$. A tolerance $\epsilon_a^{\max} \sim 10^{-2}$ is sufficient for the largest time step $\Delta t = 0.04$, the intermediate time step $\Delta t = 0.004$ requires a tolerance of $\epsilon_a^{\max} \sim 10^{-5}$, and the smallest time step $\Delta t = 0.0004$ requires $\epsilon_a^{\max} \sim 10^{-9}$. This approach avoids solving the nonlinear system of equations to a higher accuracy than required by the time integration method for a given $\Delta t$. 
The relative tolerance for the GMRES solver is specified as $\tilde{\epsilon}_r = 10^{-4}$, and the absolute tolerance is specified as $\tilde{\epsilon}_a = \epsilon_a^{\max}/10$ for all the simulations. 
Figure~\ref{fig:case1_errdt} shows that all the methods implemented converge at their theoretical orders. 
The errors for a given method are shown at all CFL numbers for which the method is stable. The explicit RK methods are unstable beyond a collisional CFL of $\sigma_C \sim 0.5$ that corresponds to a Vlasov CFL of $\sigma_V \sim 0.04$ since they are constrained by the collisional time scale. The ARK methods allow stable time steps comparable to the Vlasov time scales; the maximum Vlasov CFL is $\sigma_V \sim 1.1$, corresponding to a collisional CFL of $\sigma_C \sim 12$.

The computational cost of the ARK methods are compared with that of the explicit RK methods for the solutions shown in Figure~\ref{fig:case1_errdt}. Figure~\ref{fig:case1_errfunccount} shows the $L_2$ norm of the truncation error ${\bf f}_\epsilon$ as a function of the number of calls to the collision operator $\hat{\mathcal{C}}\left(\bar{\bf f}\right)$. It is defined as
\begin{align}
N_{\rm C} = N_T \times n_s \label{eqn:nc_rk}
\end{align}
for an explicit RK time integrator, where $N_T = t_f/\Delta t$ is the total number of time steps, and $n_s$ is the number of stages in the RK methods. Since we use the JFNK approach to solve the nonlinear system of equations in the ARK time integrators, it is defined as
\begin{align}
N_{\rm C} = N_T \times n_s + N_{\rm Newton} + N_{\rm GMRES}\label{eqn:nc_ark}
\end{align}
for the ARK methods, where $N_{\rm Newton}$ is the total number of Newton iterations and $N_{\rm GMRES}$ is the total number of preconditioned GMRES iterations. The time step $\Delta t$ (and consequently, the collisional and Vlasov CFL numbers) increases as one moves along a curve from bottom-right to the top-left. In the region where the explicit methods are stable, they are more efficient, and this is expected because of the additional burden on the ARK methods to solve an nonlinear system of equations. However, the ARK methods allow stable solutions at a significantly lower cost at higher CFL numbers where the explicit methods are unstable. The metric shown in Figure~\ref{fig:case1_errfunccount} does not reflect the cost of assembling and solving the preconditioner. The Gauss-Seidel method is used to invert the preconditioner, and $80$ iterations are required to solve it to sufficient accuracy. Figure~\ref{fig:case1_errwalltime} shows truncation error $\|\bar{\bf f}_\epsilon\|_2$ as a function of the the wall time (in seconds), which includes the additional cost of preconditioning the system. The simulations are run on $96$ cores ($6$ nodes) with $8$ MPI ranks along $y$, $6$ ranks along $v_\parallel$, and $2$ ranks along $\mu$. A similar trend is observed, and the ARK methods allow for significantly faster stable time step. For example, the fastest stable solution with ARK2 is $\sim 6$ times faster than that with RK2a, while the fastest stable solution with ARK4 is $\sim 4$ times faster than that with RK4.

\begin{table}[t]
\caption{Computational cost with and without preconditioning for solutions obtained with the ARK4 method at a final time of $t_f=130$. The preconditioner is effective at reducing the cost of the linear solver at higher CFL numbers.}
\label{tab:case1_cost}
\begin{tabular}{lrrrrrrrr}
\hline\noalign{\smallskip}
\multicolumn{1}{c}{$\Delta t$} & \multicolumn{1}{c}{$\sigma_{\rm V}$} & \multicolumn{1}{c}{$\sigma_{\rm C}$} & \multicolumn{2}{c}{$N_{\rm GMRES}$} & \multicolumn{2}{c}{$N_{\rm C}$} & \multicolumn{2}{c}{Wall time (seconds)}  \\
& & & \multicolumn{1}{c}{No PC} & \multicolumn{1}{c}{With PC} & \multicolumn{1}{c}{No PC} & \multicolumn{1}{c}{With PC} & \multicolumn{1}{c}{No PC} & \multicolumn{1}{c}{With PC} \\
\noalign{\smallskip}\hline\noalign{\smallskip}
\multicolumn{9}{l}{ARK4} \\
$0.01$ & $0.2$ & $ 2.4$ & $180,568$ & $188,974$ & $388,408$ & $397,023$ & $1.4\times10^5$ & $1.5\times10^5$ \\
$0.025$& $0.6$ & $ 6.1$ & $ 73,734$ & $ 59,096$ & $156,935$ & $142,297$ & $5.7\times10^4$ & $5.3\times10^4$ \\
$0.04$ & $0.9$ & $ 9.7$ & $ 54,321$ & $ 25,852$ & $103,801$ & $ 75,249$ & $3.7\times10^4$ & $2.7\times10^4$ \\
$0.05$ & $1.1$ & $12.1$ & $ 49,905$ & $ 21,755$ & $ 89,544$ & $ 61,298$ & $3.3\times10^4$ & $2.3\times10^4$ \\
\noalign{\smallskip}\hline\noalign{\smallskip}
\multicolumn{9}{l}{RK4} \\
$0.002$& $0.04$ & $ 0.50$ & \multicolumn{2}{c}{---} & \multicolumn{2}{c}{$260,000$} & \multicolumn{2}{c}{$1.1\times10^5$} \\
\noalign{\smallskip}\hline
\end{tabular}
\end{table}

\begin{figure}[t]
\begin{center}
\subfigure[Temperature vs. time]{\includegraphics[width=0.49\textwidth]{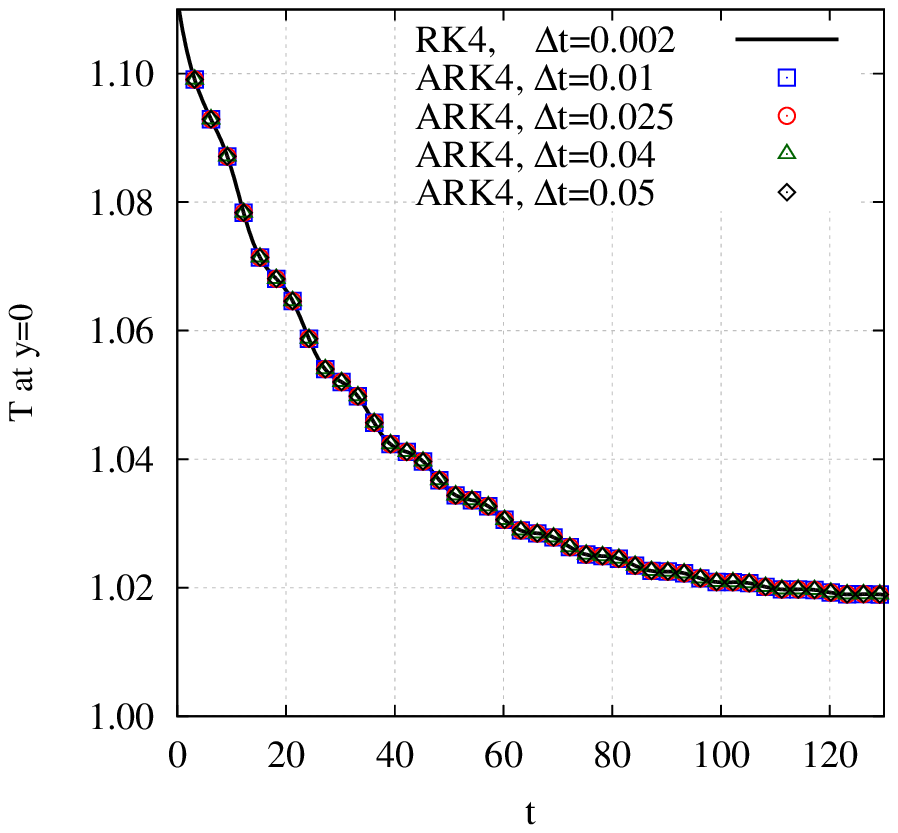}\label{fig:case1_temp_comp}}
\subfigure[Difference with respect to RK4 solution]{\includegraphics[width=0.49\textwidth]{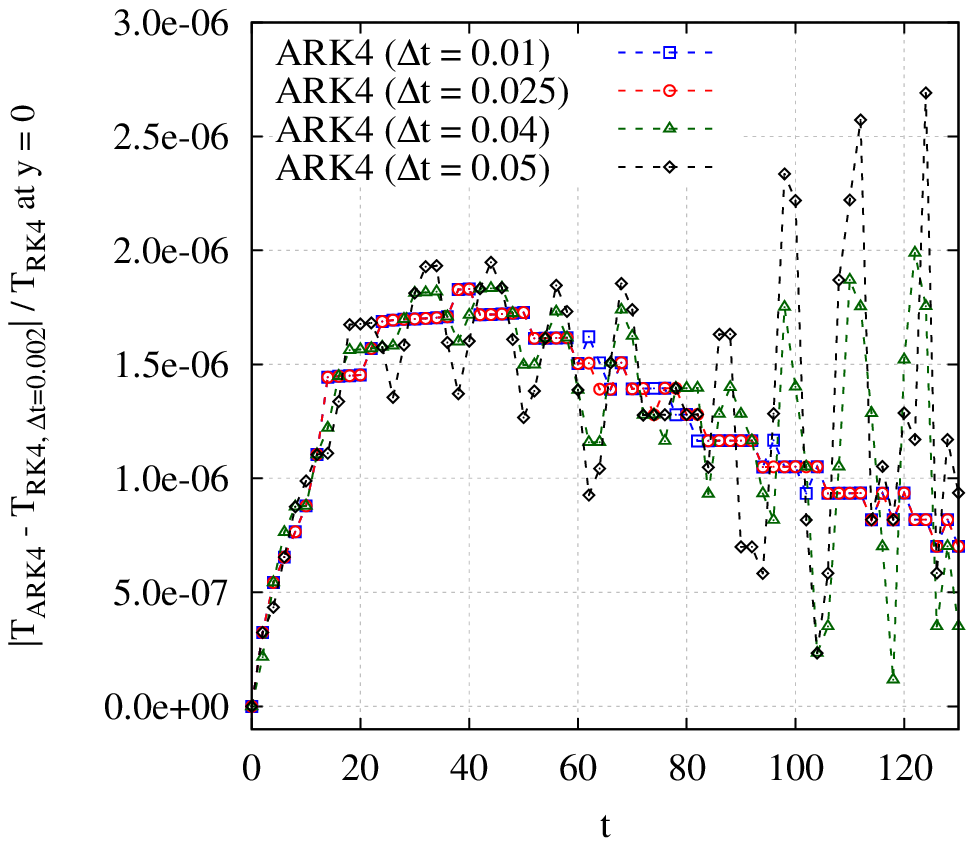}\label{fig:case1_temp_diff}}
\end{center}
\caption{(a) Temperature $T$ at $y=0$ as a function of time for the solutions reported in Table~\ref{tab:case1_cost}. The solutions obtained with the semi-implicit ARK4 method agree well with that obtained with the explicit RK4 method. (b) The normalized difference between the ARK4 and RK4 solutions.}
\label{fig:case1_temp_comp2}
\end{figure}

The performance of the preconditioner is assessed for the ARK4 method. Table~\ref{tab:case1_cost} shows the number of GMRES iterations ($N_{\rm GMRES}$), number of function calls to $\hat{\mathcal{C}}$ ($N_{\rm C}$), and the wall time for solutions obtained with various time steps $\Delta t$ at final time of $t_f = 130$. The simulations are run with the preconditioner (``With PC") and without it (``No PC") on $192$ cores ($12$ nodes) with $8$ MPI ranks along $y$, $6$ ranks along $v_\parallel$, and $4$ ranks along $\mu$. As a reference, the number of function calls and the wall time are reported for the explicit RK4 with the largest stable time step $\Delta t = 0.002$; this represents the fastest stable solution with the RK4 method. The preconditioner is ineffective at the lower CFL numbers considered; at $\sigma_{\rm C} = 2.4$, the number of preconditioned GMRES iterations is slightly larger than the number of unpreconditioned GMRES iterations, while at $\sigma_{\rm C} = 6.1$, the number of preconditioned iterations is slightly less than the number of unpreconditioned iterations. At the higher CFL numbers $\sigma_{\rm C} = 9.7, 12.1$, the preconditioner reduces the number of GMRES iterations by factors of $\sim 2.1$ and $\sim 2.3$, respectively. The number of function calls $N_{\rm C}$ includes the calls from the Newton solver and the time integrator, and therefore, the effect of the preconditioner is less pronounced when considering this metric. At the highest stable time step for the ARK4 method ($\Delta t = 0.05$), the number of function calls for the preconditioned simulation is $\sim 1.5$ times smaller than that for the unpreconditioned simulation. In terms of the wall time, the preconditioned simulation is $\sim 1.4$ times faster than the unpreconditioned simulation, and this shows that the construction and inversion of the preconditioner is relatively cheap even though $80$ Gauss-Seidel iterations are required. Overall, this test case does not result in a very ill-conditioned system; at the highest collisional CFL number considered, the unpreconditioned simulation requires only $\sim 4$ GMRES iterations on average ($N_{\rm GMRES}/(N_T\times n_s^{\rm imp})$, where $n_s^{\rm imp}=5$ is the number of implicit stages for ARK4 and the nonlinearity is sufficiently weak that only one Newton iteration is observed for each implicit stage). Figure~\ref{fig:case1_temp_comp} shows the evolution of the temperature $T$ at $y=0$ for the simulations in Table~\ref{tab:case1_cost}. The solutions obtained with the semi-implicit approach at high collisional CFL numbers agree well with the solution obtained with the explicit RK4 method with a time step constrained by the collisional time scale. Figure~\ref{fig:case1_temp_diff} shows the normalized difference between the ARK4 solutions ($T_{\rm ARK4}$) and the fastest RK4 solution ($T_{\rm RK4, \Delta t = 0.002}$). The errors are larger for larger time steps, and they grow with time as expected; the global (in time) error at a given time is an accumulation of the local truncation error for all preceding time steps.

\begin{figure}[t]
\subfigure[Initial density $n_0\left(x\right)$]{\includegraphics[width=0.49\textwidth]{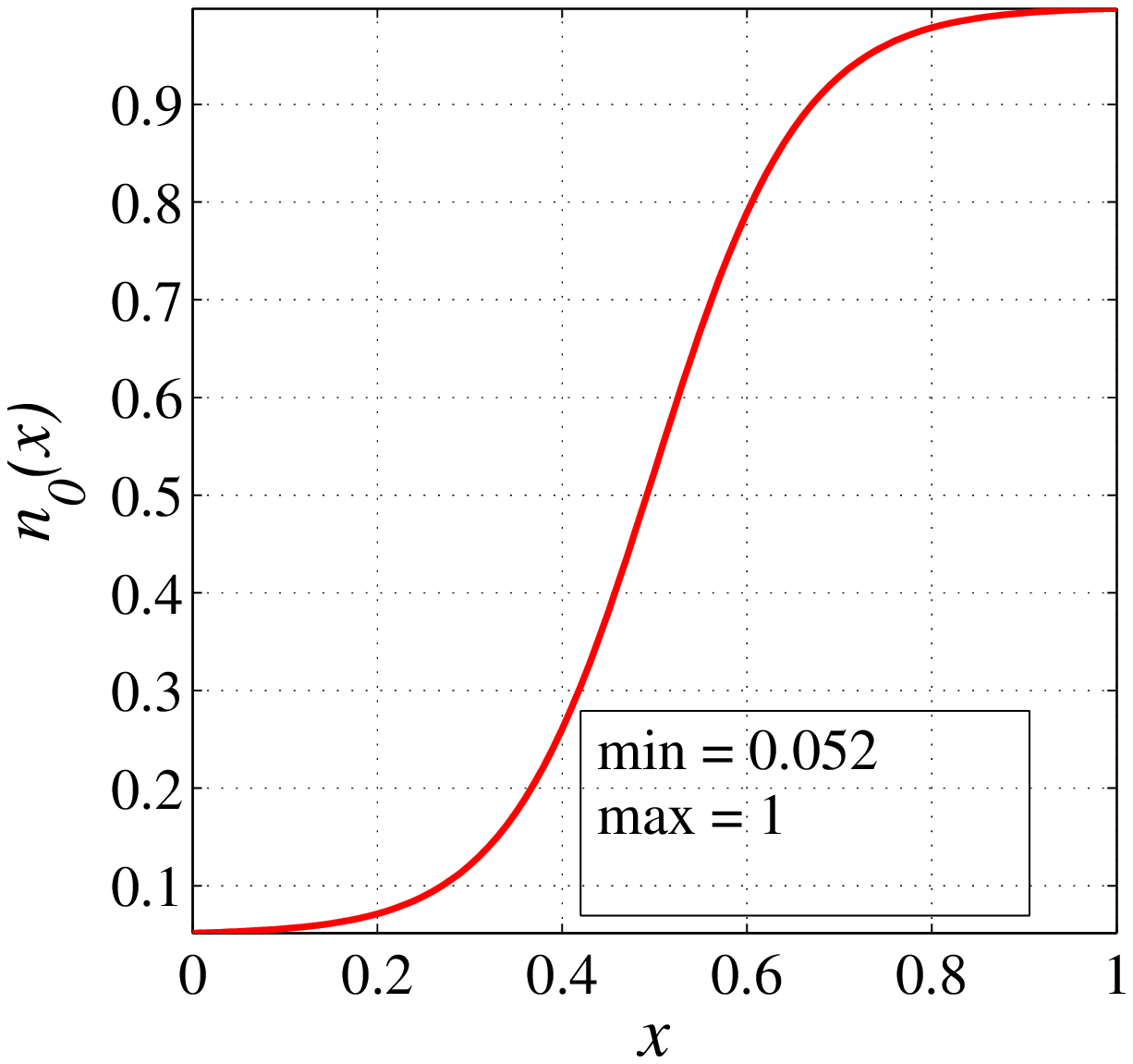}\label{fig:case2_density}}
\subfigure[Ratio of mean free path to length scale $k_\parallel \lambda$]{\includegraphics[width=0.49\textwidth]{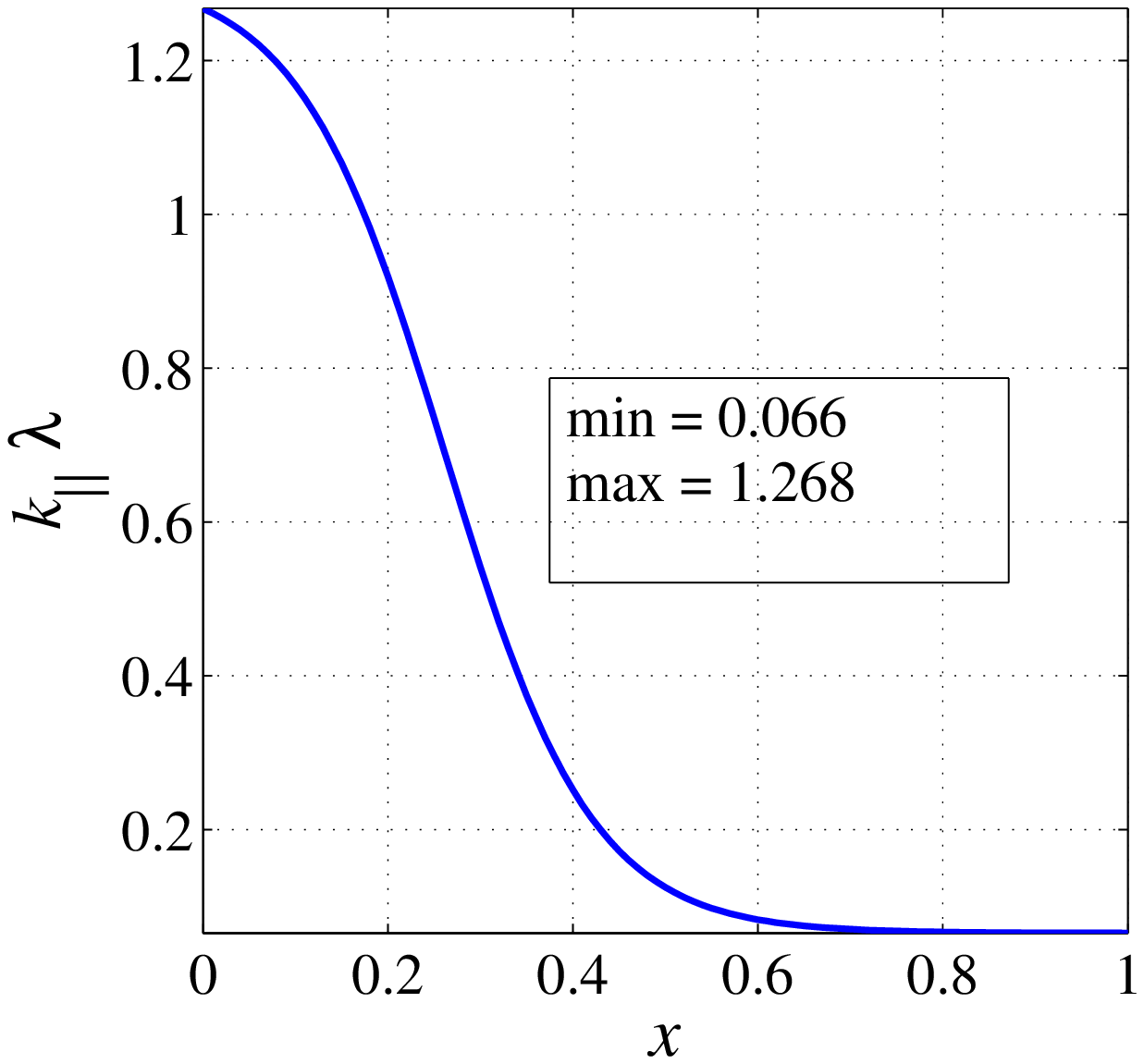}\label{fig:case2_k_par_lambda}}
\caption{The initial density variation along $x$ for the varying collisionality case and the consequent collisionality variation. The plasma is weakly collisional for $x \rightarrow 0$ while it is highly collisional for $x \rightarrow 1$.}
\label{fig:case2_vary}
\end{figure}

\subsection{Case 2: Varying Collisionality}
\label{sec:case2}

\begin{figure}[b]
\subfigure[]{\includegraphics[width=0.49\textwidth]{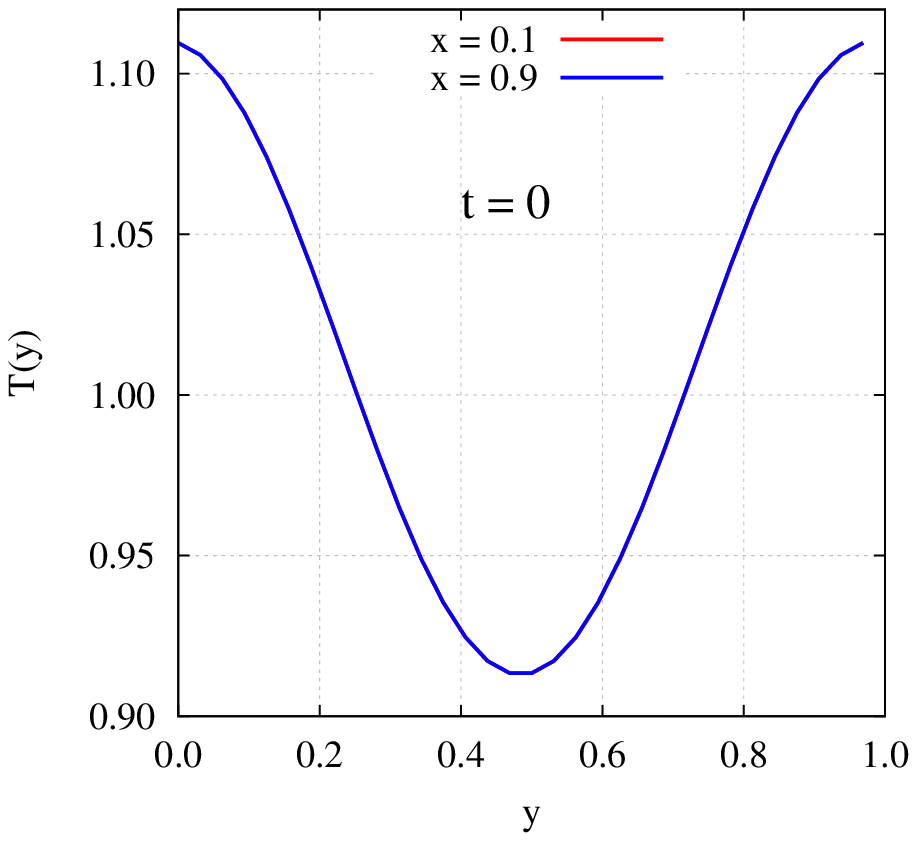}\label{fig:case2_temperature_000}}
\subfigure[]{\includegraphics[width=0.49\textwidth]{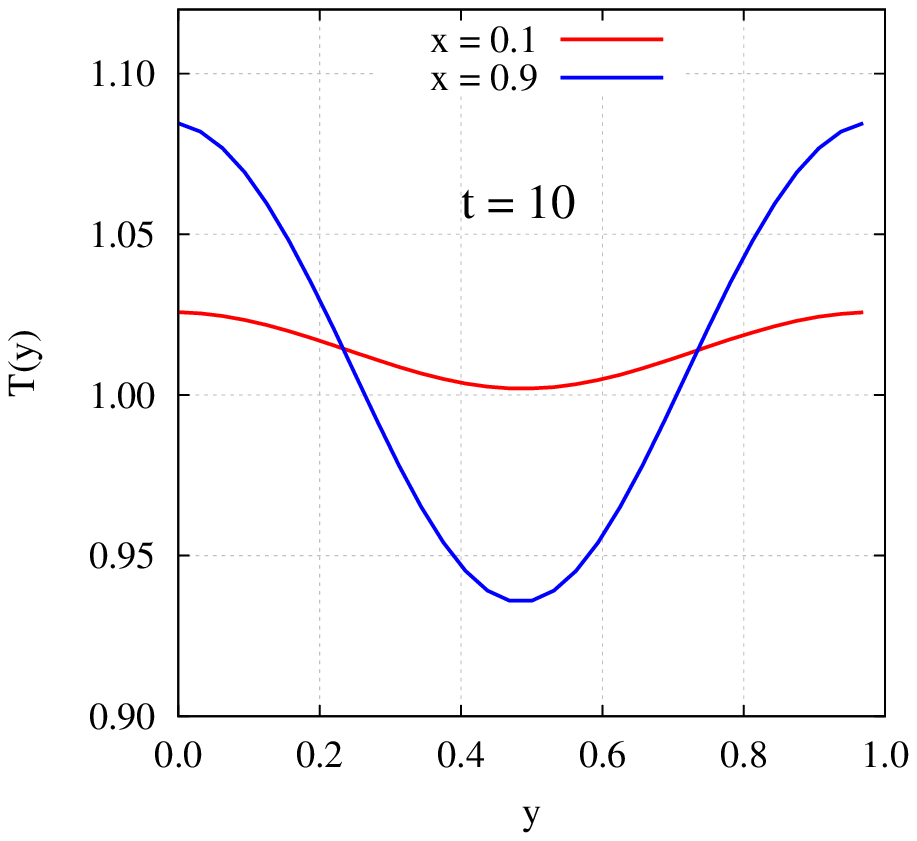}\label{fig:case2_temperature_002}}
\subfigure[]{\includegraphics[width=0.49\textwidth]{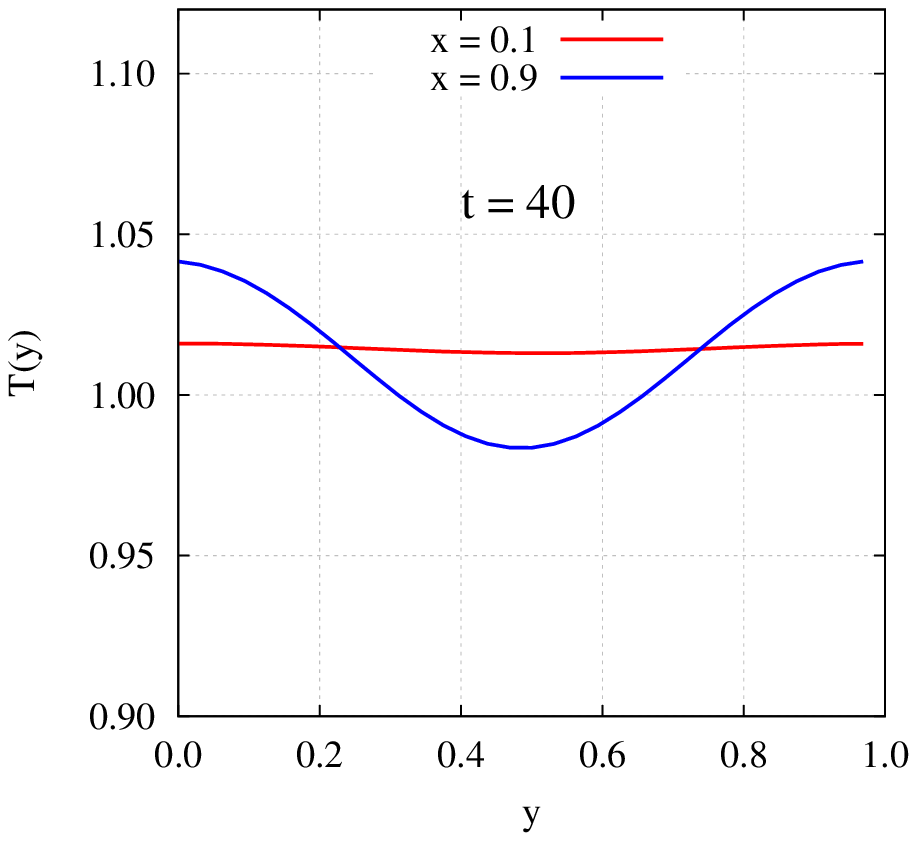}\label{fig:case2_temperature_004}}
\subfigure[]{\includegraphics[width=0.49\textwidth]{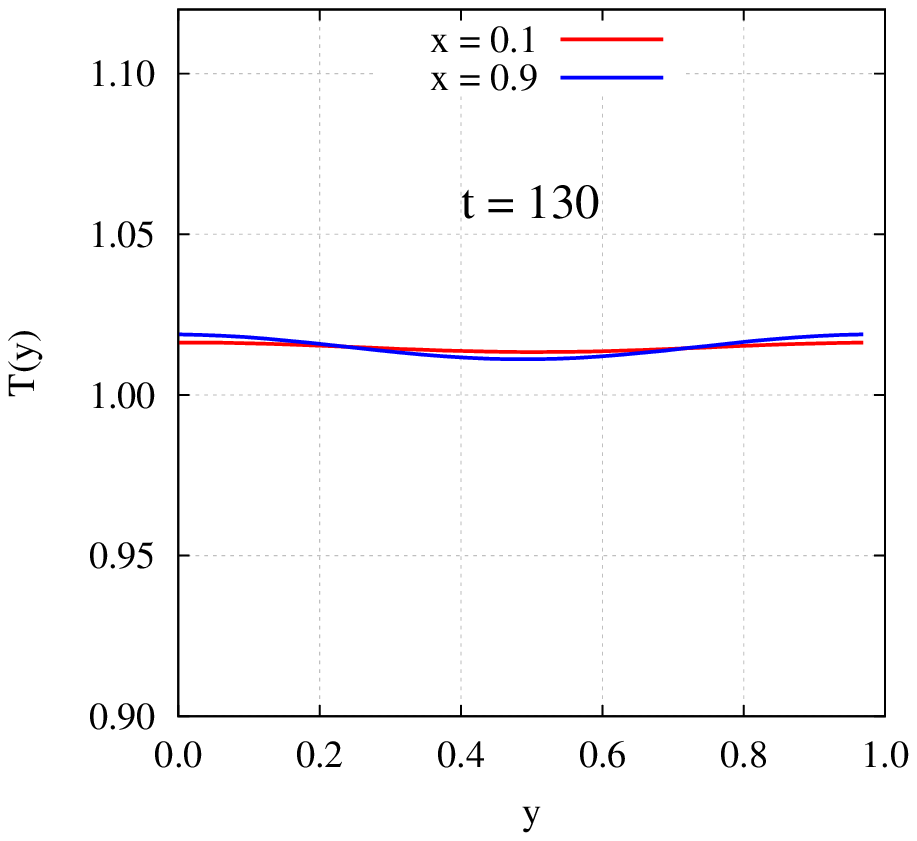}\label{fig:case2_temperature_006}}
\caption{Cross-sectional temperature $T\left(y\right)$ at $x=0.1,0.9$ at four different solution times. The temperature equilibrates faster in the low-collisionality region ($x=0.1$) than in the high-collisionality region ($x=0.9$).}
\label{fig:case2_temperature}
\end{figure}

This case is similar to the previous case, except that a significant variation in the density, and therefore, the collisionality, is introduced along $x$. The initial density is specified as
\begin{align}
n\left(x,y,t=0\right) \equiv n_0\left(x\right) = 1 - \delta + \delta \tanh{\left(2\pi x - \pi\right)},
\end{align}
where $\delta = 0.475$. The problem setup is identical to the previous case in all other aspects. Figure~\ref{fig:case2_vary} shows the initial density and the ratio of the mean free path to the characteristic length scale $k_\parallel \lambda$. The mean free path is comparable to the characteristic length scale near $x=0$, and the plasma is weakly collisional. On the other hand, the mean free path is much smaller than the length scale near $x=1$, and the plasma is strongly collisional. The collisional time scales are thus comparable to the Vlasov time scales as $x\rightarrow 0$ but are much smaller as $x \rightarrow 1$. This variation in the collisionality is representative of the radial variation in the tokamak edge region~\cite{porteretal2000}. The plasma conditions as $x\rightarrow 1$ correspond to the previous test case. This case is solved on a grid with $\left(N_x, N_y, N_{v_\parallel}, N_\mu\right) = \left(32, 32, 36, 24\right)$ points in the subsequent discussions.

\begin{figure}[t]
\begin{center}
\includegraphics[width=0.7\textwidth]{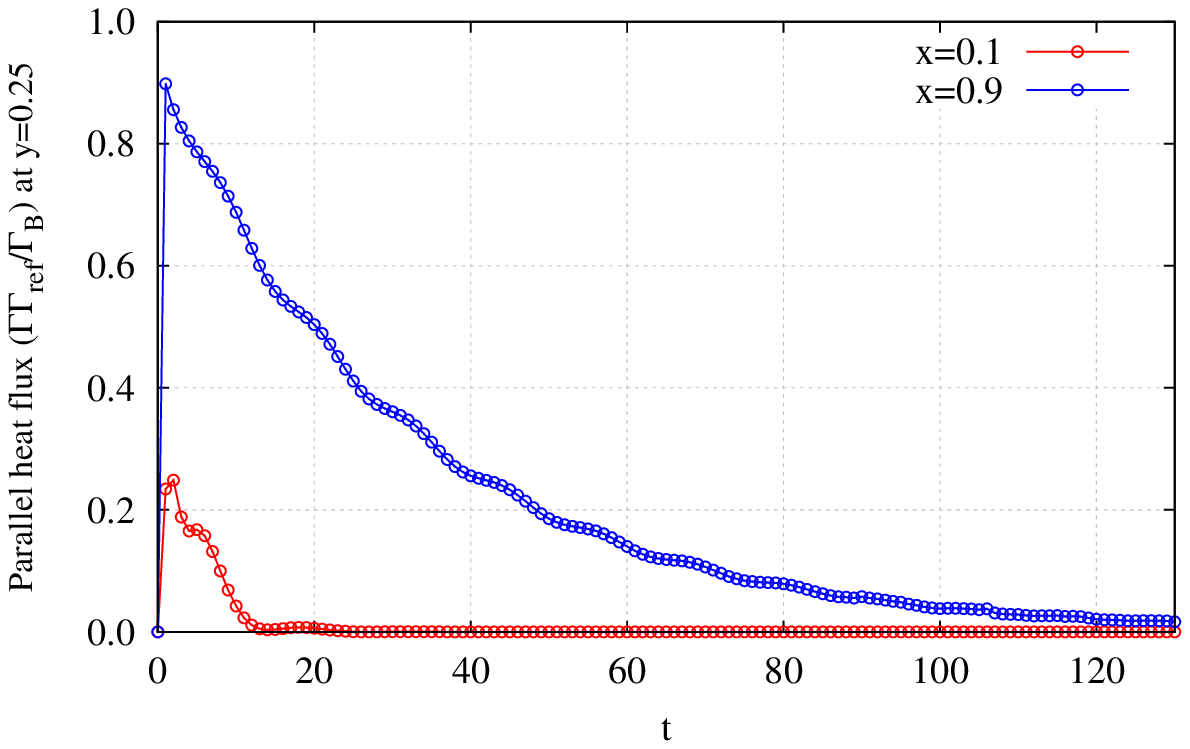}
\end{center}
\caption{The time evolution of the parallel heat flux at $y=0.25$ and $x=0.1,0.9$. The dynamics are qualitatively similar to those in Fig.~\ref{fig:case1_phf}. In the low collisionality region ($x=0.1$), thermal equilibration is faster and the heat flux decays to zero faster than in the high collisionality region ($x=0.9$).}
\label{fig:case2_phf}
\end{figure}

\begin{figure}[t]
\begin{center}
\subfigure[Error vs. Vlasov and collisional CFL numbers]{\includegraphics[width=0.6\textwidth] {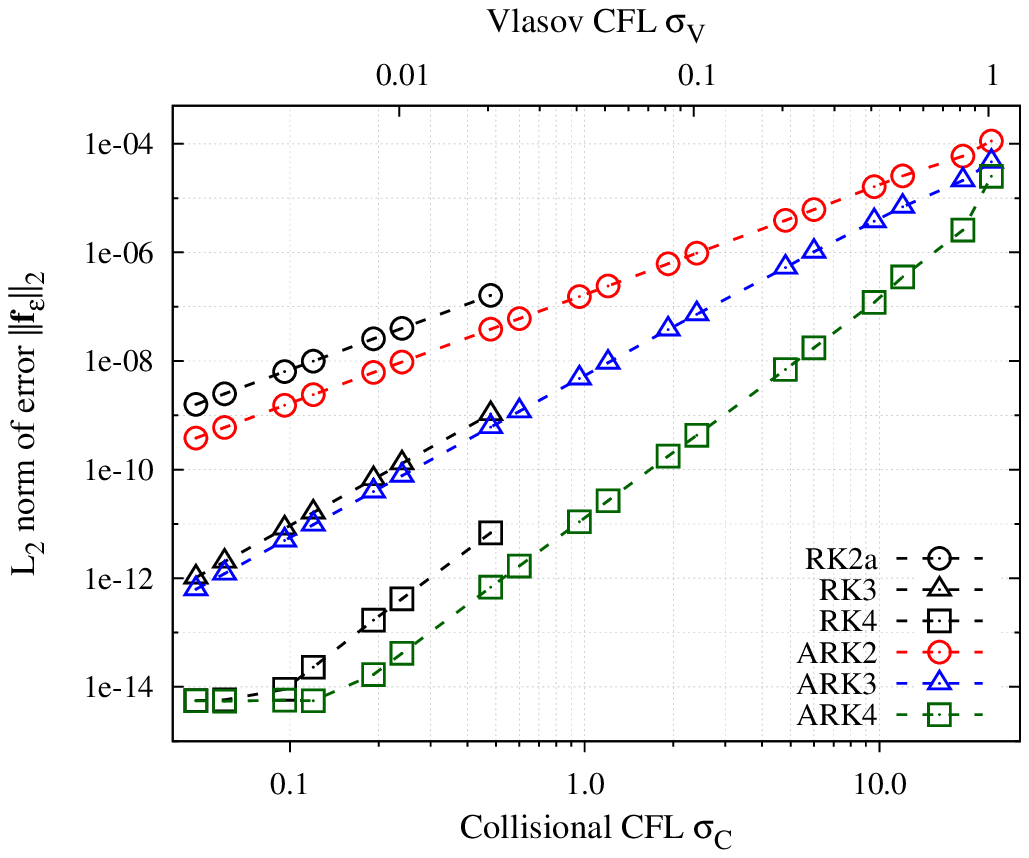}\label{fig:case2_errdt}}
\subfigure[Error vs. number of function calls]{\includegraphics[width=0.49\textwidth]{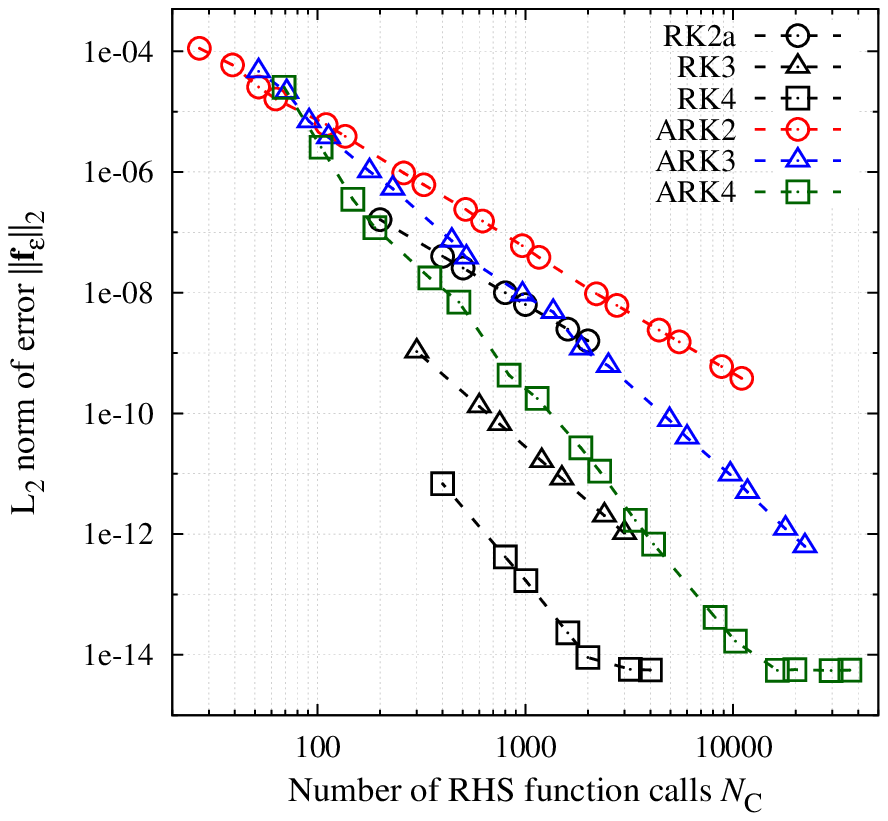}\label{fig:case2_errfunccount}}
\subfigure[Error vs. wall time]{\includegraphics[width=0.49\textwidth]{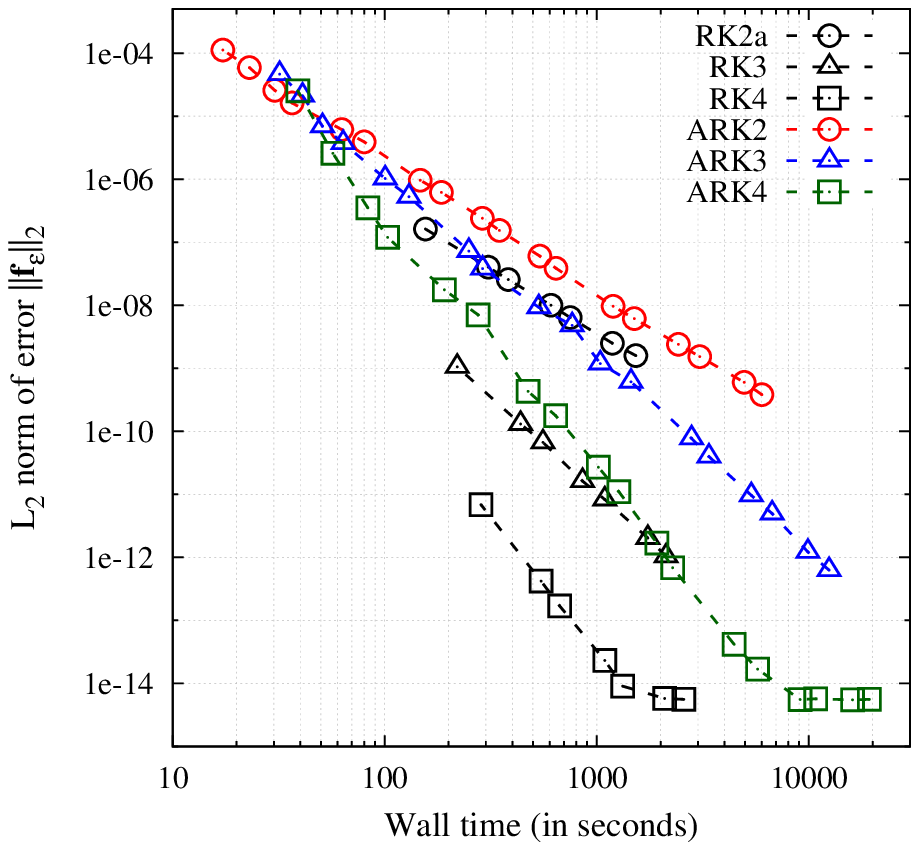}\label{fig:case2_errwalltime}}
\end{center}
\caption{Convergence and work-precision: (a) The error as a function of the Vlasov and collisional CFL numbers for explicit RK and semi-implicit ARK methods. All the methods converge at their theoretical orders. (b) The error as a function of the number of calls to the collision operator $\hat{\mathcal{C}}$, and (c) the error as a function of the wall time in seconds. The ARK methods allow significantly faster stable solutions.}
\label{fig:case2_conv}
\end{figure}

Figure~\ref{fig:case2_temperature} shows the cross-sectional temperature along $y$ at $x=0.1$ (weakly collisional) and $x=0.9$ (strongly collisional). The solutions are obtained with the ARK4 method at a time step of $\Delta t = 0.1$, corresponding to a Vlasov CFL number of $\sigma_{\rm V} \approx 1.1$ and a collisional CFL number of $\sigma_{\rm C} \approx 26$. The simulation is run until a final time of $t_f = 130$ on $384$ cores ($24$ nodes) with $4$ MPI ranks along $x$, $y$, and $\mu$ and $6$ MPI ranks along $v_\parallel$. The dynamics are similar to the previous test case; however, the varying collisionality results in varying thermal equilibration and density diffusion rates. Near $x=0$, the collisional and Vlasov time scales are comparable while near $x=1$, the collisional time scale is much smaller than the Vlasov time scale. In the region where the plasma is weakly collisional ($x\rightarrow 0$), the temperature equilibrates much faster while the equilibration is much slower in the region where the plasma is strongly collisional ($x\rightarrow 1$). Figure~\ref{fig:case2_phf} shows the evolution of the normalized parallel heat flux ($\Gamma \Gamma_{\rm ref}/\Gamma_B$) at $y=0.25$ and $x=0.1,0.9$. At $x=0.1$, the heat flux decays to zero much faster than at $x=0.9$ due to the faster equilibration.

Figure~\ref{fig:case2_errdt} shows the $L_2$ norm of the time integration error as a function of the Vlasov and collisional CFL numbers defined by~(\ref{eqn:cfl}) for solutions at a final time of $t_f = 0.2$. The errors are computed with respect to a reference solution obtained with the RK4 method and a time step of $\Delta t = 2\times10^{-5}$, corresponding to a Vlasov CFL number of $\sigma_{\rm V} \approx 2.2\times10^{-4}$ and a collisional CFL number of $\sigma_{\rm C} \approx 4.7\times10^{-3}$. Thus, the fastest collisional time scale is $\sim 21$ times faster than the Vlasov time scale for this test case. The relative and absolute tolerances for the Newton and GMRES solvers are specified in the same way as described for the previous test case. All the methods considered converge at their theoretical orders of convergence. The maximum stable time step for the explicit RK methods correspond to the collisional CFL number of $\sigma_{\rm C} \approx 0.5$ because they are constrained by the collisional time scales. The semi-implicit methods are stable until a collisional CFL of $\sigma_{\rm C} \approx 25$ that corresponds to a Vlasov CFL of $\sigma_{\rm V} \approx 1.1$; thus, they are constrained by the Vlasov time scale. 

Figure~\ref{fig:case2_errfunccount} shows the error as a function of the number of calls ($N_{\rm C}$) to the collision operator $\hat{\mathcal{C}}\left(\bar{\bf f}\right)$ defined through~(\ref{eqn:nc_rk}) and~(\ref{eqn:nc_ark}) for the explicit RK and semi-implicit ARK methods, respectively. The ARK methods allow significantly less expensive stable solutions; for example, the fastest stable solution with ARK2 is $\sim 7$ times less expensive than that with RK2a, and the fastest stable solution with ARK4 is $\sim 6$ times less expensive than that with RK4. Figure~\ref{fig:case2_errwalltime} shows the error as a function of the wall time. The simulations are run on $192$ cores with $4$ MPI ranks along $x$ and $y$, $6$ ranks along $v_\parallel$, and $2$ ranks along $\mu$. The wall times include the cost of assembling and inverting the preconditioning matrix ($80$ Gauss-Seidel iterations), and similar trends are observed. Thus, the additional cost of the preconditioner is insignificant despite $80$ Gauss-Seidel iterations being required to solve it to sufficient accuracy.

\begin{table}[t]
\caption{A comparison of the computational cost with and without preconditioning of the fastest stable solution with the ARK4 method at a final time of $t_f=130$, as well as the fastest stable solution with RK4. The preconditioner is effective at reducing the cost of the linear solve.}
\label{tab:case2_cost}
\begin{tabular}{lrrrrrrrr}
\hline\noalign{\smallskip}
\multicolumn{1}{c}{$\Delta t$} & \multicolumn{1}{c}{$\sigma_{\rm V}$} & \multicolumn{1}{c}{$\sigma_{\rm C}$} & \multicolumn{2}{c}{$N_{\rm GMRES}$} & \multicolumn{2}{c}{$N_{\rm C}$} & \multicolumn{2}{c}{Wall time (seconds)}  \\
& & & \multicolumn{1}{c}{No PC} & \multicolumn{1}{c}{With PC} & \multicolumn{1}{c}{No PC} & \multicolumn{1}{c}{With PC} & \multicolumn{1}{c}{No PC} & \multicolumn{1}{c}{With PC} \\
\noalign{\smallskip}\hline\noalign{\smallskip}
\multicolumn{9}{l}{ARK4} \\
$0.1$ & $1.1$ & $25.8$ & $ 33,131$ & $ 11,535$ & $ 52,551$ & $ 30,852$ & $2.5\times10^4$ & $1.6\times10^4$ \\
\noalign{\smallskip}\hline\noalign{\smallskip}
\multicolumn{9}{l}{RK4} \\
$0.002$& $0.02$ & $ 0.52$ & \multicolumn{2}{c}{---} & \multicolumn{2}{c}{$260,000$} & \multicolumn{2}{c}{$1.7\times10^5$} \\
\noalign{\smallskip}\hline
\end{tabular}
\end{table}

\begin{figure}[t]
\subfigure[$x=0.1$]{\includegraphics[width=0.49\textwidth]{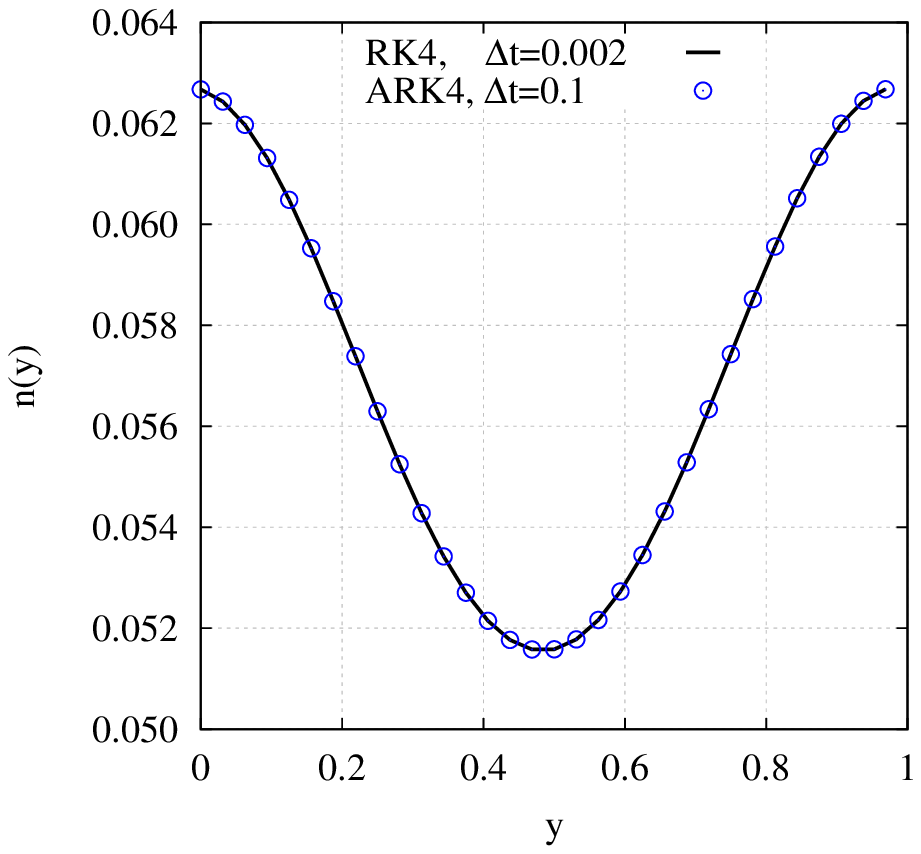}\label{fig:case2_density_x0.1_comp}}
\subfigure[$x=0.9$]{\includegraphics[width=0.49\textwidth]{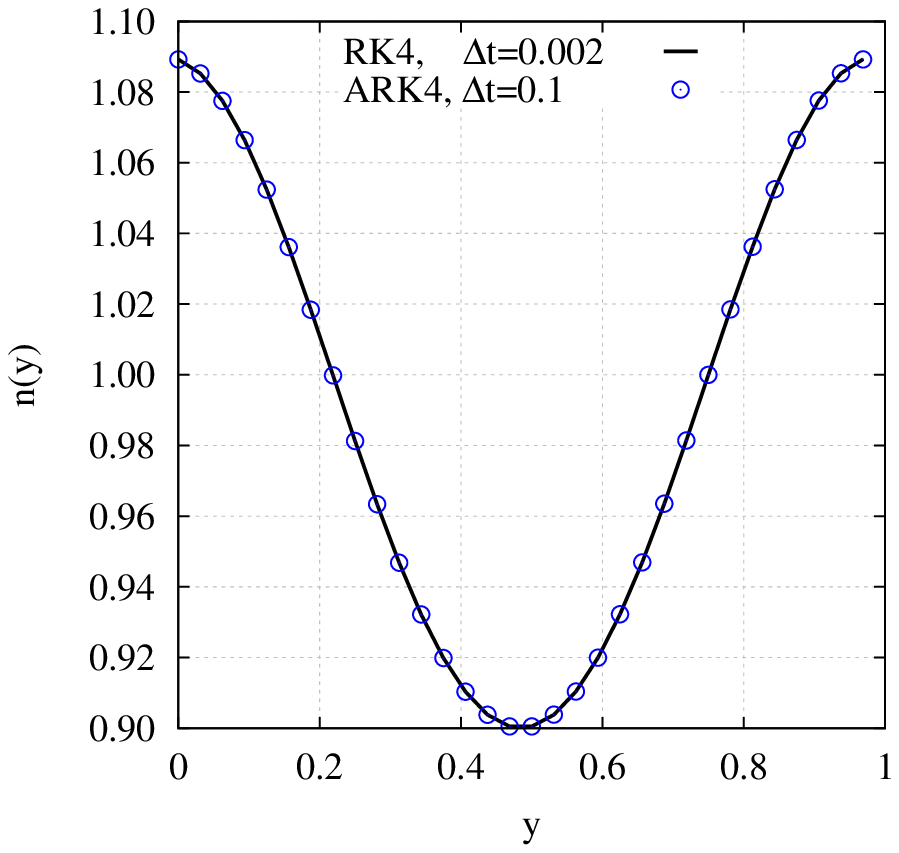}\label{fig:case2_density_x0.9_comp}}
\caption{Cross-sectional density $n\left(y\right)$ at two $x$ locations corresponding to the weakly and strongly collisional regions for the simulations in Table~\ref{tab:case2_cost} at the final simulation time of $t_f=130$. The solutions agree well with each other.}
\label{fig:case2_density_comp}
\end{figure}

Table~\ref{tab:case2_cost} compares the computational cost of the fastest stable solution with the ARK4 method, both with and without the preconditioner, as well as the fastest stable RK4 solution. The preconditioner reduces the total number of GMRES iterations by a factor of $\sim 3$. The unpreconditioned simulation requires an average of $\sim 5$ GMRES iterations in each implicit stage, while the preconditioned simulation requires an average of $\sim 1.8$ GMRES iterations. The nonlinearity was sufficiently weak that the Newton solver was observed to converge in one iteration for these simulations. The total number of calls to the collision operator ($N_{\rm C}$), which includes calls from the time integrator and the Newton solver, is reduced by a factor of $\sim 1.7$ by the preconditioner, while the total wall time is reduced by $\sim 1.6$. The difference in these two factors reflects the cost of inverting the preconditioning matrix with $80$ Gauss-Seidel iterations. The preconditioned ARK4 solution is $\sim 11$ times faster than the fastest stable RK4 solution in terms of the wall time. These solutions are obtained on $384$ cores ($24$ nodes) with $4$ MPI ranks along $x$, $y$, and $\mu$, and $6$ ranks along $v_\parallel$. Figure~\ref{fig:case2_density_comp} shows the cross-sectional density $n\left(y\right)$ at $x=0.1$ and $x=0.9$ for the two simulations in Table~\ref{tab:case2_cost} at the final time of $t_f = 130$. The initially constant density developes a non-constant profile to balance the specified electrostatic potential. The solution obtained with the semi-implicit ARK4 method agrees well with that obtained with the explicit RK4 method.

\section{Conclusion}
\label{sec:summary}
This paper describes a semi-implicit time integration algorithm for the Vlasov-Fokker-Planck equations with the motivating application area being tokamak edge plasma simulations. The dynamics in the edge region of a tokamak fusion reactor are characterized by disparate length and time scales. The hot, weakly-collisional plasma near the core requires a kinetic model; however, near the cold edge, the strong collisionality renders the kinetic model numerically stiff. Explicit time integration methods are thus too expensive for such applications. In this paper, we consider as our governing equations the drift-kinetic Vlasov equation, which is the long-wavelength limit of the gyrokinetic VFP equation, with the Fokker-Planck collision model in the Rosenbluth form. 

We implement high-order, semi-implicit Additive Runge-Kutta methods where the Vlasov term is integrated explicitly in time while the Fokker-Planck term is integrated implicitly. The semi-implicit algorithms are applied to two test problems, where the ratio of the collisional time scales to the Vlasov time scales are representative of the tokamak edge region. We show that the semi-implicit time integrators allow time steps that respect the Vlasov time scales and are larger than the maximum time steps allowed by explicit time integrators, which respect the collisional time scales. Accuracy and convergence tests of the methods show that the semi-implicit methods converge at their theoretical orders. We also assess the computational cost of these methods and compare the relative cost of the semi-implicit solutions with those obtained with explicit Runge-Kutta methods. Overall, the IMEX approach results in significantly faster stable solutions. Explicit methods are more efficient when seeking solutions with high accuracy. However, the IMEX approach allows faster solutions that are less accurate (due to a higher truncation error from a larger time step size) but well-resolved; explicit methods cannot yield solutions at a comparable cost even allowing for lower accuracy since their time step sizes are constrained.

Two deficiencies of this paper are areas of active research and will be reported in future publications. A better preconditioning approach is needed, both in terms of constructing the preconditioning matrix as an approximation to the true Jacobian and implementing an efficient method to solve it. The use of fixed velocity grids restricts us to applications without severe temperature gradients, and thus, our current implementation cannot solve for true tokamak conditions where the temperature drops significantly from the hot core to the cold edge. This limitation will be addressed by using velocity renormalization.

\begin{appendix}
\section{Computation of Rosenbluth Potentials}
\label{app:rosenbluth}

The Rosenbluth potentials are related to the distribution function by Poisson equations, given by (\ref{eqn:rosenbluth}), that are defined on an infinite velocity domain. The algorithm implemented in COGENT to solve them on a finite numerical domain was introduced in~\cite{dorf_etal_2014} and is summarized briefly in this section. The numerical velocity domain is defined as $\Omega_{\bf v} \equiv \left[-v_{\parallel,{\rm max} },v_{\parallel,{\rm max} }\right] \times \left[0,\mu_{\rm max}\right]$, while the infinite velocity domain is $\Omega_{{\bf v},\infty} \equiv \left[-\infty,\infty\right] \times \left[0,\infty\right]$. The Green's function method is used to compute the boundary values as
\begin{subequations}\label{eqn:rosenbluthbc}
\begin{align}
\varphi\left({\bf v}_{\partial\Omega_{\bf v}}\right) &= \frac{1}{4\pi} \int\displaylimits_{\Omega_{{\bf v},\infty}} \frac{f\left({\bf v}'\right)}{\left| {\bf v}_{\partial\Omega_{\bf v}} - {\bf v}'  \right|} d{\bf v}',\label{eqn:rosenbluthbc1} \\
\varrho\left({\bf v}_{\partial\Omega_{\bf v}}\right) &= \frac{1}{8\pi} \int\displaylimits_{\Omega_{{\bf v},\infty}} \left| {\bf v}_{\partial\Omega_{\bf v}} - {\bf v}'  \right| f\left({\bf v}'\right) d{\bf v}',\label{eqn:rosenbluthbc2}
\end{align}
\end{subequations}
where ${\bf v}_{\partial\Omega_{\bf v}}$ is the velocity vector at the computational domain boundary $\partial\Omega_{\bf v}$. Direct evaluation of (\ref{eqn:rosenbluthbc}) is expensive and an asymptotic method is used. The Green's function is expanded as~\cite{jackson}
\begin{align}
\frac{1}{\left|{\bf v} - {\bf v}'\right|} = 4\pi \sum_{l=0}^{\infty} \sum_{m=-l}^{l} \frac{1}{2l+1} \frac{v_{<}^l}{v_{>}^{l+1}}Y_{lm}^*\left(\theta',\psi'\right)Y_{lm}\left(\theta,\psi\right),
\end{align}
where $Y_{lm}$ is the spherical harmonic function, $\theta = \arccos\left(v_\parallel/v\right)$ is the pitch angle, $\psi$ is the gyro-angle, $v_{>} = \max{\left( \left|{\bf v}\right|, \left|{\bf v}'\right| \right)}$, and $v_{<} = \min{\left( \left|{\bf v}\right|, \left|{\bf v}'\right| \right)}$. Therefore,
\begin{align}
\varphi\left({\bf v}_{\partial\Omega_{\bf v}}\right) = -\frac{1}{4\pi} \sum_{l=0}^{\infty} \frac{h_l}{\left|{\bf v}_{\partial\Omega_{\bf v}}\right|^{l+1}} P_l\left(\cos{\theta}\right);\ \  h_l = \frac{2\pi B}{m} \int\displaylimits_{\Omega_{\bf v}} f\left(v_\parallel,\mu\right)v^l P_l\left(\cos{\theta}\right) dv_\parallel d\mu,\label{eqn:rosenbluth_bc_1}
\end{align}
where $P_l$ denotes the Legendre polynomials, $v = \left|{\bf v}\right|$ is the particle speed, and
\begin{align}
f\left(v_\parallel, \mu: v > \min{\left|{\bf v}_{\partial\Omega_{\bf v}}\right|}\right) = 0.
\end{align}
The second Rosenbluth potential $\varrho$ is expressed on the domain boundary in terms of the first Rosenbluth potential $\varphi$ as
\begin{align}
\varrho\left({\bf v}_{\partial\Omega_{\bf v}}\right) &= \frac{1}{4\pi} \int\displaylimits_{\Omega_{\infty}} \frac{\varphi\left({\bf v}'\right)}{\left| {\bf v}_{\partial\Omega_{\bf v}} - {\bf v}'  \right|} d{\bf v}';\label{eqn:rosenbluthbc2_alt}.
\end{align}
Though this is similar to (\ref{eqn:rosenbluthbc1}), the above analysis does not apply because $\varphi\left(v_\parallel, \mu: v > \min{\left|{\bf v}_{\partial\Omega_{\bf v}}\right|}\right) \ne 0$.
Equation~(\ref{eqn:rosenbluthbc2_alt}) is decomposed as
\begin{align}
\frac{1}{4\pi} \int\displaylimits_{\Omega_{\infty}} \frac{\varphi\left({\bf v}'\right)}{\left| {\bf v}_{\partial\Omega_{\bf v}} - {\bf v}'  \right|} d{\bf v}'
=
\int\displaylimits_{0}^{\min{\left|{\bf v}_{\partial\Omega_{\bf v}}\right|}} v^2 dv \int\displaylimits_{0}^{\pi} d\theta \int\displaylimits_{0}^{2\pi} \frac{\hat{\varphi}\left({\bf v}'\right)}{\left| {\bf v}_{\partial\Omega_{\bf v}} - {\bf v}'  \right|} d\psi 
+
\int\displaylimits_{0}^{\min{\left|{\bf v}_{\partial\Omega_{\bf v}}\right|}} v^2 dv \int\displaylimits_{0}^{\pi} d\theta \int\displaylimits_{0}^{2\pi} \frac{\tilde{\varphi}\left({\bf v}'\right)}{\left| {\bf v}_{\partial\Omega_{\bf v}} - {\bf v}'  \right|} d\psi,
\end{align}
where $\hat{\varphi}\left({\bf v}\right)$ is the numerical solution to Eq.~(\ref{eqn:rosenbluth1}) and 
\begin{align}
\tilde{\varphi}\left({\bf v}\right) = -\frac{1}{4\pi} \sum_{l=0}^{\infty} \frac{h_l}{\left|{\bf v}\right|^{l+1}} P_l\left(\cos{\theta}\right).
\end{align}
Therefore,
\begin{align}
\int\displaylimits_{0}^{\min{\left|{\bf v}_{\partial\Omega_{\bf v}}\right|}} v^2 dv \int\displaylimits_{0}^{\pi} d\theta \int\displaylimits_{0}^{2\pi} \frac{\hat{\varphi}\left({\bf v}'\right)}{\left| {\bf v}_{\partial\Omega_{\bf v}} - {\bf v}'  \right|} d\psi = \sum_{l=0}^{\infty} \frac{g_l}{\left|{\bf v}_{\partial\Omega_{\bf v}}\right|^{l+1}}P_l\left(\cos{\theta}\right);\ \ g_l = \frac{2\pi B}{m} \int\displaylimits_{\Omega_{\bf v}} \hat{\hat{\varphi}}\left(v_\parallel,\mu\right)v^l P_l\left(\cos{\theta}\right) dv_\parallel d\mu,\label{eqn:gl_int}
\end{align}
where $\hat{\hat{\varphi}}\left({\bf v}\right) = \hat{\varphi}\left({\bf v}\right), v \le \min{\left|{\bf v}_{\partial\Omega_{\bf v}}\right|}$, $\hat{\hat{\varphi}}\left({\bf v}\right) = 0, v > \min{\left|{\bf v}_{\partial\Omega_{\bf v}}\right|}$, and
\begin{align}
\int\displaylimits_{0}^{\min{\left|{\bf v}_{\partial\Omega_{\bf v}}\right|}} v^2 dv \int\displaylimits_{0}^{\pi} d\theta \int\displaylimits_{0}^{2\pi} \frac{\tilde{\varphi}\left({\bf v}'\right)}{\left| {\bf v}_{\partial\Omega_{\bf v}} - {\bf v}'  \right|} d\psi
=
-\sum_{l=0}^{\infty} \frac{h_l P_l\left(\cos{\theta}\right)}{2l+1} \left( \frac{1}{2\left|{\bf v}_{\partial\Omega_{\bf v}}\right|^{l-1}}  - \frac{1}{2\left|{\bf v}_{\partial\Omega_{\bf v}}\right|^{l+1}} + \frac{1}{\left(2l-1\right)\left|{\bf v}_{\partial\Omega_{\bf v}}\right|^{l-1}} \right).
\end{align}
The second Rosenbluth potential is thus expressed at the computational domain boundary as
\begin{align}
\varrho\left({\bf v}_{\partial\Omega_{\bf v}}\right) = -\frac{1}{4\pi} \sum_{l=0}^{\infty} \left[ \frac{g_l}{\left|{\bf v}_{\partial\Omega_{\bf v}}\right|^{l+1}} - \frac{h_l}{2\left(2l+1\right)} \left( \frac{2l+1}{2l-1}\frac{}{\left|{\bf v}_{\partial\Omega_{\bf v}}\right|^{l-1}} - \frac{\left\{\min{\left|{\bf v}_{\partial\Omega_{\bf v}}\right|}\right\}^2}{\left|{\bf v}_{\partial\Omega_{\bf v}}\right|^{l+1}} \right) \right] P_l\left(\cos{\theta}\right).\label{eqn:rosenbluth_bc_2}
\end{align}
Equations~(\ref{eqn:rosenbluth_bc_1}) and (\ref{eqn:rosenbluth_bc_2}) serve as the boundary conditions for (\ref{eqn:rosenbluth}) on the finite numerical domain $\Omega_v$. Cut-cell issues, arising in evaluating (\ref{eqn:gl_int}), are solved by linear interpolations~\cite{dorf_etal_2014}, and therefore, the current implementation uses second-order central finite differences to discretize the derivatives in Eq.~(\ref{eqn:rosenbluth}). The fourth-order implementation will be investigated in the future. We use the conjugate gradient method with a structured multigrid preconditioner from the {\it hypre} library~\cite{hypre} to solve the discretized system of equations.

\end{appendix}






\bibliographystyle{spmpsci}
\bibliography{plasma,time,atmos,weno}

\begin{thebibliography}{10}
\providecommand{\url}[1]{{#1}}
\providecommand{\urlprefix}{URL }
\expandafter\ifx\csname urlstyle\endcsname\relax
  \providecommand{\doi}[1]{DOI~\discretionary{}{}{}#1}\else
  \providecommand{\doi}{DOI~\discretionary{}{}{}\begingroup
  \urlstyle{rm}\Url}\fi

\bibitem{ascherruuthspiteri}
Ascher, U.M., Ruuth, S.J., Spiteri, R.J.: Implicit-explicit {R}unge-{K}utta
  methods for time-dependent partial differential equations.
\newblock Applied Numerical Mathematics \textbf{25}(2-3), 151--167 (1997).
\newblock \doi{10.1016/S0168-9274(97)00056-1}

\bibitem{banksetal2016}
Banks, J.W., Brunner, S., Berger, R.L., Tran, T.M.: Vlasov simulations of
  electron-ion collision effects on damping of electron plasma waves.
\newblock Physics of Plasmas \textbf{23}(3), 032108 (2016).
\newblock \doi{10.1063/1.4943194}

\bibitem{bellicandy2012}
Belli, E.A., Candy, J.: Full linearized {F}okker--{P}lanck collisions in
  neoclassical transport simulations.
\newblock Plasma Physics and Controlled Fusion \textbf{54}(1), 015,015 (2012).
\newblock \doi{10.1088/0741-3335/54/1/015015}

\bibitem{berezinetal1987}
Berezin, Y., Khudick, V., Pekker, M.: Conservative finite-difference schemes
  for the {F}okker-{P}lanck equation not violating the law of an increasing
  entropy.
\newblock Journal of Computational Physics \textbf{69}(1), 163--174 (1987).
\newblock \doi{10.1016/0021-9991(87)90160-4}

\bibitem{braginskii}
Braginskii, S.I.: Transport processes in a plasma.
\newblock Reviews of Plasma Physics \textbf{1}, 205 (1965)

\bibitem{brunneretal2010}
Brunner, S., Tran, T., Hittinger, J.: Numerical implementation of the
  non-linear {L}andau collision operator for {E}ulerian {V}lasov simulations.
  part {I}: Computation of the {R}osenbluth potentials.
\newblock Tech. Rep. LLNL-SR-459135, Lawrence Livermore National Laboratory,
  Livermore, CA (2010)

\bibitem{buetcordier1998}
Buet, C., Cordier, S.: Conservative and entropy decaying numerical scheme for
  the isotropic {F}okker--{P}lanck--{L}andau equation.
\newblock Journal of Computational Physics \textbf{145}(1), 228--245 (1998).
\newblock \doi{10.1006/jcph.1998.6015}

\bibitem{buetetal1997}
Buet, C., Cordier, S., Degond, P., Lemou, M.: Fast algorithms for numerical,
  conservative, and entropy approximations of the {F}okker--{P}lanck--{L}andau
  equation.
\newblock Journal of Computational Physics \textbf{133}(2), 310--322 (1997).
\newblock \doi{10.1006/jcph.1997.5669}

\bibitem{butcher2003}
Butcher, J.: Numerical Methods for Ordinary Differential Equations.
\newblock Wiley (2003)

\bibitem{candywaltz2003}
Candy, J., Waltz, R.E.: Anomalous transport scaling in the {DIII--D} tokamak
  matched by supercomputer simulation.
\newblock Phys. Rev. Lett. \textbf{91}, 045,001 (2003).
\newblock \doi{10.1103/PhysRevLett.91.045001}

\bibitem{casanovaetal1991}
Casanova, M., Larroche, O., Matte, J.P.: Kinetic simulation of a collisional
  shock wave in a plasma.
\newblock Phys. Rev. Lett. \textbf{67}, 2143--2146 (1991).
\newblock \doi{10.1103/PhysRevLett.67.2143}

\bibitem{chaconetal2000_1}
Chac{\'o}n, L., Barnes, D.C., Knoll, D.A., Miley, G.H.: An implicit
  energy-conservative 2{D} {F}okker--{P}lanck algorithm: {I}. difference
  scheme.
\newblock Journal of Computational Physics \textbf{157}(2), 618--653 (2000).
\newblock \doi{10.1006/jcph.1999.6394}

\bibitem{chaconetal2000_2}
Chac{\'o}n, L., Barnes, D.C., Knoll, D.A., Miley, G.H.: An implicit
  energy-conservative 2{D} {F}okker--{P}lanck algorithm: {II}. {J}acobian-free
  {N}ewton--{K}rylov solver.
\newblock Journal of Computational Physics \textbf{157}(2), 654--682 (2000).
\newblock \doi{10.1006/jcph.1999.6395}

\bibitem{changku2008}
Chang, C.S., Ku, S.: Spontaneous rotation sources in a quiescent tokamak edge
  plasma.
\newblock Physics of Plasmas \textbf{15}(6), 062510 (2008).
\newblock \doi{10.1063/1.2937116}

\bibitem{chang1970}
Chang, J., Cooper, G.: A practical difference scheme for {F}okker-{P}lanck
  equations.
\newblock Journal of Computational Physics \textbf{6}(1), 1--16 (1970).
\newblock \doi{10.1016/0021-9991(70)90001-X}

\bibitem{chenchacon2014}
Chen, G., Chac{\'o}n, L.: An energy- and charge-conserving, nonlinearly
  implicit, electromagnetic 1d-3v {V}lasov--{D}arwin particle-in-cell
  algorithm.
\newblock Computer Physics Communications \textbf{185}(10), 2391--2402 (2014).
\newblock \doi{10.1016/j.cpc.2014.05.010}

\bibitem{chenchacon2015}
Chen, G., Chac{\'o}n, L.: A multi-dimensional, energy- and charge-conserving,
  nonlinearly implicit, electromagnetic {V}lasov--{D}arwin particle-in-cell
  algorithm.
\newblock Computer Physics Communications \textbf{197}, 73--87 (2015).
\newblock \doi{10.1016/j.cpc.2015.08.008}

\bibitem{chenetal2011}
Chen, G., Chac{\'o}n, L., Barnes, D.: An energy- and charge-conserving,
  implicit, electrostatic particle-in-cell algorithm.
\newblock Journal of Computational Physics \textbf{230}(18), 7018--7036 (2011).
\newblock \doi{10.1016/j.jcp.2011.05.031}

\bibitem{chengknorr1976}
Cheng, C., Knorr, G.: The integration of the {V}lasov equation in configuration
  space.
\newblock Journal of Computational Physics \textbf{22}(3), 330--351 (1976).
\newblock \doi{10.1016/0021-9991(76)90053-X}

\bibitem{cohenxu2008}
Cohen, R.H., Xu, X.Q.: Progress in kinetic simulation of edge plasmas.
\newblock Contributions to Plasma Physics \textbf{48}(1-3), 212--223 (2008).
\newblock \doi{10.1002/ctpp.200810038}

\bibitem{collelaetal2011}
Colella, P., Dorr, M., Hittinger, J., Martin, D.: High--order, finite--volume
  methods in mapped coordinates.
\newblock Journal of Computational Physics \textbf{230}(8), 2952--2976 (2011).
\newblock \doi{10.1016/j.jcp.2010.12.044}

\bibitem{crouseillesetal2009}
Crouseilles, N., Respaud, T., Sonnendr{\"u}cker, E.: A forward
  semi-{L}agrangian method for the numerical solution of the {V}lasov equation.
\newblock Computer Physics Communications \textbf{180}(10), 1730--1745 (2009).
\newblock \doi{10.1016/j.cpc.2009.04.024}

\bibitem{dennisschnabel}
Dennis, J., Schnabel, R.: Numerical Methods for Unconstrained Optimization and
  Nonlinear Equations.
\newblock Society for Industrial and Applied Mathematics (1996).
\newblock \doi{10.1137/1.9781611971200}

\bibitem{dorf_etal_2012}
Dorf, M.A., Cohen, R.H., Compton, J.C., Dorr, M., Rognlien, T.D., Angus, J.,
  Krasheninnikov, S., Colella, P., Martin, D., McCorquodale, P.: Progress with
  the {COGENT} edge kinetic code: Collision operator options.
\newblock Contributions to Plasma Physics \textbf{52}(5-6), 518--522 (2012).
\newblock \doi{10.1002/ctpp.201210042}

\bibitem{dorf_etal_2014}
Dorf, M.A., Cohen, R.H., Dorr, M., Hittinger, J., Rognlien, T.D.: Progress with
  the {COGENT} edge kinetic code: Implementing the {F}okker-{P}lanck collision
  operator.
\newblock Contributions to Plasma Physics \textbf{54}(4-6), 517--523 (2014).
\newblock \doi{10.1002/ctpp.201410023}

\bibitem{dorf_etal_2013}
Dorf, M.A., Cohen, R.H., Dorr, M., Rognlien, T., Hittinger, J., Compton, J.,
  Colella, P., Martin, D., McCorquodale, P.: Simulation of neoclassical
  transport with the continuum gyrokinetic code {COGENT}.
\newblock Physics of Plasmas \textbf{20}(1), 012513 (2013).
\newblock \doi{10.1063/1.4776712}

\bibitem{durranblossey}
Durran, D.R., Blossey, P.N.: Implicit--explicit multistep methods for
  fast-wave--slow-wave problems.
\newblock Monthly Weather Review \textbf{140}(4), 1307--1325 (2012).
\newblock \doi{10.1175/MWR-D-11-00088.1}

\bibitem{epperlein1994}
Epperlein, E.: Implicit and conservative difference scheme for the
  {F}okker-{P}lanck equation.
\newblock Journal of Computational Physics \textbf{112}(2), 291 -- 297 (1994).
\newblock \doi{10.1006/jcph.1994.1101}

\bibitem{epperleinetal1988}
Epperlein, E.M., Rickard, G.J., Bell, A.R.: Two-dimensional nonlocal electron
  transport in laser-produced plasmas.
\newblock Physical Review Letters \textbf{61}, 2453--2456 (1988).
\newblock \doi{10.1103/PhysRevLett.61.2453}

\bibitem{hypre}
Falgout, R.D., Yang, U.M.: hypre: A Library of High Performance
  Preconditioners, pp. 632--641.
\newblock Springer Berlin Heidelberg, Berlin, Heidelberg (2002).
\newblock \doi{10.1007/3-540-47789-6_66}

\bibitem{filbet2002}
Filbet, F., Pareschi, L.: A numerical method for the accurate solution of the
  {F}okker--{P}lanck--{L}andau equation in the nonhomogeneous case.
\newblock Journal of Computational Physics \textbf{179}(1), 1--26 (2002).
\newblock \doi{10.1006/jcph.2002.7010}

\bibitem{filbetetal2001}
Filbet, F., Sonnendr{\"u}cker, E., Bertrand, P.: Conservative numerical schemes
  for the {V}lasov equation.
\newblock Journal of Computational Physics \textbf{172}(1), 166--187 (2001).
\newblock \doi{10.1006/jcph.2001.6818}

\bibitem{ghoshconstaSISC2016}
Ghosh, D., Constantinescu, E.M.: Semi-implicit time integration of atmospheric
  flows with characteristic-based flux partitioning.
\newblock SIAM Journal on Scientific Computing \textbf{38}(3), A1848--A1875
  (2016).
\newblock \doi{10.1137/15M1044369}

\bibitem{giraldokellyconsta2013}
Giraldo, F.X., Kelly, J.F., Constantinescu, E.: Implicit-explicit formulations
  of a three-dimensional nonhydrostatic unified model of the atmosphere
  ({NUMA}).
\newblock SIAM Journal on Scientific Computing \textbf{35}(5), B1162--B1194
  (2013).
\newblock \doi{10.1137/120876034}

\bibitem{giraldorestellilauter2010}
Giraldo, F.X., Restelli, M., L{\"a}uter, M.: Semi-implicit formulations of the
  {N}avier-{S}tokes equations: Application to nonhydrostatic atmospheric
  modeling.
\newblock SIAM Journal on Scientific Computing \textbf{32}(6), 3394--3425
  (2010).
\newblock \doi{10.1137/090775889}

\bibitem{gysela}
Grandgirard, V., Brunetti, M., Bertrand, P., Besse, N., Garbet, X., Ghendrih,
  P., Manfredi, G., Sarazin, Y., Sauter, O., Sonnendr{\"u}cker, E., Vaclavik,
  J., Villard, L.: A drift-kinetic semi-{L}agrangian {4D} code for ion
  turbulence simulation.
\newblock Journal of Computational Physics \textbf{217}(2), 395--423 (2006).
\newblock \doi{10.1016/j.jcp.2006.01.023}

\bibitem{hahm1996}
Hahm, T.S.: Nonlinear gyrokinetic equations for turbulence in core transport
  barriers.
\newblock Physics of Plasmas \textbf{3}(12), 4658--4664 (1996).
\newblock \doi{10.1063/1.872034}

\bibitem{heikkinenetal2001}
Heikkinen, J., Kiviniemi, T., Kurki-Suonio, T., Peeters, A., Sipil{\"a}, S.:
  Particle simulation of the neoclassical plasmas.
\newblock Journal of Computational Physics \textbf{173}(2), 527---548 (2001).
\newblock \doi{10.1006/jcph.2001.6891}

\bibitem{heikkinenetal2006}
Heikkinen, J.A., Henriksson, S., Janhunen, S., Kiviniemi, T.P., Ogando, F.:
  Gyrokinetic simulation of particle and heat transport in the presence of wide
  orbits and strong profile variations in the edge plasma.
\newblock Contributions to Plasma Physics \textbf{46}(7-9), 490--495 (2006).
\newblock \doi{10.1002/ctpp.200610035}

\bibitem{nrlformulary}
Huba, J.D.: {NRL} plasma formulary.
\newblock Tech. rep., Naval Research Laboratory, Washington, DC (2016)

\bibitem{idomuraetal2008_1}
Idomura, Y., Ida, M., Kano, T., Aiba, N., Tokuda, S.: Conservative global
  gyrokinetic toroidal full-f five-dimensional {V}lasov simulation.
\newblock Computer Physics Communications \textbf{179}(6), 391--403 (2008).
\newblock \doi{10.1016/j.cpc.2008.04.005}

\bibitem{idomuraetal2008_2}
Idomura, Y., Ida, M., Tokuda, S.: Conservative gyrokinetic {V}lasov simulation.
\newblock Communications in Nonlinear Science and Numerical Simulation
  \textbf{13}(1), 227--233 (2008).
\newblock \doi{10.1016/j.cnsns.2007.05.015}.
\newblock Vlasovia 2006: The Second International Workshop on the Theory and
  Applications of the Vlasov Equation

\bibitem{jackson}
Jackson, J.D.: Classical electrodynamics, 3rd ed. edn.
\newblock Wiley, New York, {NY} (1999)

\bibitem{jacobshesthaven2009}
Jacobs, G., Hesthaven, J.: Implicit--explicit time integration of a high-order
  particle-in-cell method with hyperbolic divergence cleaning.
\newblock Computer Physics Communications \textbf{180}(10), 1760--1767 (2009).
\newblock \doi{10.1016/j.cpc.2009.05.020}

\bibitem{james1977}
James, R.: The solution of {P}oisson's equation for isolated source
  distributions.
\newblock Journal of Computational Physics \textbf{25}(2), 71--93 (1977).
\newblock \doi{10.1016/0021-9991(77)90013-4}

\bibitem{jiangshu}
Jiang, G.S., Shu, C.W.: Efficient implementation of weighted {ENO} schemes.
\newblock Journal of Computational Physics \textbf{126}(1), 202--228 (1996).
\newblock \doi{10.1006/jcph.1996.0130}

\bibitem{kadiogluetal2010}
Kadioglu, S.Y., Knoll, D.A., Lowrie, R.B., Rauenzahn, R.M.: A second order
  self-consistent {IMEX} method for radiation hydrodynamics.
\newblock Journal of Computational Physics \textbf{229}(22), 8313--8332 (2010).
\newblock \doi{10.1016/j.jcp.2010.07.019}

\bibitem{kennedycarpenter}
Kennedy, C.A., Carpenter, M.H.: Additive {R}unge-{K}utta schemes for
  convection-diffusion-reaction equations.
\newblock Applied Numerical Mathematics \textbf{44}(1-2), 139--181 (2003).
\newblock \doi{10.1016/S0168-9274(02)00138-1}

\bibitem{kho1985}
Kho, T.H.: Relaxation of a system of charged particles.
\newblock Physics Review A \textbf{32}, 666--669 (1985).
\newblock \doi{10.1103/PhysRevA.32.666}

\bibitem{kinghambell2004}
Kingham, R., Bell, A.: An implicit {V}lasov--{F}okker--{P}lanck code to model
  non-local electron transport in 2--{D} with magnetic fields.
\newblock Journal of Computational Physics \textbf{194}(1), 1--34 (2004).
\newblock \doi{10.1016/j.jcp.2003.08.017}

\bibitem{knollkeyes2004}
Knoll, D., Keyes, D.: {J}acobian-free {N}ewton--{K}rylov methods: a survey of
  approaches and applications.
\newblock Journal of Computational Physics \textbf{193}(2), 357--397 (2004).
\newblock \doi{10.1016/j.jcp.2003.08.010}

\bibitem{kumarmishra2012}
Kumar, H., Mishra, S.: Entropy stable numerical schemes for two-fluid plasma
  equations.
\newblock Journal of Scientific Computing \textbf{52}(2), 401--425 (2012).
\newblock \doi{10.1007/s10915-011-9554-7}

\bibitem{landau1937}
Landau, L.D.: The kinetic equation in the case of {C}oulomb interaction.
\newblock Zh. Eksper. i Teoret. Fiz. \textbf{7}(2), 203--209 (1937)

\bibitem{landremanernst2013}
Landreman, M., Ernst, D.R.: New velocity-space discretization for continuum
  kinetic calculations and {F}okker--{P}lanck collisions.
\newblock Journal of Computational Physics \textbf{243}, 130--150 (2013).
\newblock \doi{10.1016/j.jcp.2013.02.041}

\bibitem{larroche1993}
Larroche, O.: Kinetic simulation of a plasma collision experiment.
\newblock Physics of Fluids B \textbf{5}(8), 2816--2840 (1993).
\newblock \doi{10.1063/1.860670}

\bibitem{larroche2003}
Larroche, O.: Kinetic simulations of fuel ion transport in {ICF} target
  implosions.
\newblock The European Physical Journal D - Atomic, Molecular, Optical and
  Plasma Physics \textbf{27}(2), 131--146 (2003).
\newblock \doi{10.1140/epjd/e2003-00251-1}

\bibitem{larsenetal1985}
Larsen, E., Levermore, C., Pomraning, G., Sanderson, J.: Discretization methods
  for one-dimensional {F}okker-{P}lanck operators.
\newblock Journal of Computational Physics \textbf{61}(3), 359 -- 390 (1985).
\newblock \doi{10.1016/0021-9991(85)90070-1}

\bibitem{lee1983}
Lee, W.W.: Gyrokinetic approach in particle simulation.
\newblock Physics of Fluids \textbf{26}(2), 556--562 (1983).
\newblock \doi{10.1063/1.864140}

\bibitem{lemou1998}
Lemou, M.: Multipole expansions for the {F}okker-{P}lanck-{L}andau operator.
\newblock Numerische Mathematik \textbf{78}(4), 597--618 (1998).
\newblock \doi{10.1007/s002110050327}

\bibitem{lemou2004}
Lemou, M., Mieussens, L.: Fast implicit schemes for the
  {F}okker--{P}lanck--{L}andau equation.
\newblock Comptes Rendus Mathematique \textbf{338}(10), 809--814 (2004).
\newblock \doi{10.1016/j.crma.2004.03.010}

\bibitem{lemou2005}
Lemou, M., Mieussens, L.: Implicit schemes for the {F}okker--{P}lanck--{L}andau
  equation.
\newblock SIAM Journal on Scientific Computing \textbf{27}(3), 809--830 (2005).
\newblock \doi{10.1137/040609422}

\bibitem{maeyamaetal2012}
Maeyama, S., Ishizawa, A., Watanabe, T.H., Nakajima, N., Tsuji-Iio, S.,
  Tsutsui, H.: A hybrid method of semi-{L}agrangian and additive semi-implicit
  {R}unge--{K}utta schemes for gyrokinetic {V}lasov simulations.
\newblock Computer Physics Communications \textbf{183}(9), 1986--1992 (2012).
\newblock \doi{10.1016/j.cpc.2012.04.028}

\bibitem{mccorquodaleetal2015}
McCorquodale, P., Dorr, M., Hittinger, J., Colella, P.: High--order
  finite--volume methods for hyperbolic conservation laws on mapped multiblock
  grids.
\newblock Journal of Computational Physics \textbf{288}, 181--195 (2015).
\newblock \doi{10.1016/j.jcp.2015.01.006}

\bibitem{mccoyetal1981}
McCoy, M., Mirin, A., Killeen, J.: {FPPAC}: A two-dimensional multispecies
  nonlinear {F}okker-{P}lanck package.
\newblock Computer Physics Communications \textbf{24}(1), 37--61 (1981).
\newblock \doi{10.1016/0010-4655(81)90105-3}

\bibitem{mousseauknoll1997}
Mousseau, V., Knoll, D.: Fully implicit kinetic solution of collisional
  plasmas.
\newblock Journal of Computational Physics \textbf{136}(2), 308--323 (1997).
\newblock \doi{10.1006/jcph.1997.5736}

\bibitem{pareschirusso}
Pareschi, L., Russo, G.: Implicit-explicit {R}unge-{K}utta schemes and
  applications to hyperbolic systems with relaxation.
\newblock Journal of Scientific Computing \textbf{25}(1-2), 129--155 (2005).
\newblock \doi{10.1007/BF02728986}

\bibitem{pataki2011}
Pataki, A., Greengard, L.: Fast elliptic solvers in cylindrical coordinates and
  the {C}oulomb collision operator.
\newblock Journal of Computational Physics \textbf{230}(21), 7840--7852 (2011).
\newblock \doi{10.1016/j.jcp.2011.07.005}

\bibitem{pernicewalker}
Pernice, M., Walker, H.F.: {NITSOL}: A {N}ewton iterative solver for nonlinear
  systems.
\newblock SIAM Journal on Scientific Computing \textbf{19}(1), 302--318 (1998).
\newblock \doi{10.1137/S1064827596303843}

\bibitem{porteretal2000}
Porter, G.D., Isler, R., Boedo, J., Rognlien, T.D.: Detailed comparison of
  simulated and measured plasma profiles in the scrape-off layer and edge
  plasma of {DIII-D}.
\newblock Physics of Plasmas \textbf{7}(9), 3663--3680 (2000).
\newblock \doi{10.1063/1.1286509}

\bibitem{qiuchristlieb2010}
Qiu, J.M., Christlieb, A.: A conservative high order semi-{L}agrangian {WENO}
  method for the {V}lasov equation.
\newblock Journal of Computational Physics \textbf{229}(4), 1130--1149 (2010).
\newblock \doi{10.1016/j.jcp.2009.10.016}

\bibitem{qiushu2011}
Qiu, J.M., Shu, C.W.: Positivity preserving semi-{L}agrangian discontinuous
  {G}alerkin formulation: Theoretical analysis and application to the
  {V}lasov--{P}oisson system.
\newblock Journal of Computational Physics \textbf{230}(23), 8386--8409 (2011).
\newblock \doi{10.1016/j.jcp.2011.07.018}

\bibitem{rosenbluth1957}
Rosenbluth, M.N., MacDonald, W.M., Judd, D.L.: {F}okker-{P}lanck equation for
  an inverse-square force.
\newblock The Physical Review \textbf{107}, 1--6 (1957).
\newblock \doi{10.1103/PhysRev.107.1}

\bibitem{rossmanithseal2011}
Rossmanith, J.A., Seal, D.C.: A positivity-preserving high-order
  semi-{L}agrangian discontinuous {G}alerkin scheme for the {V}lasov--{P}oisson
  equations.
\newblock Journal of Computational Physics \textbf{230}(16), 6203--6232 (2011).
\newblock \doi{10.1016/j.jcp.2011.04.018}

\bibitem{saad}
Saad, Y.: Iterative Methods for Sparse Linear Systems: Second Edition.
\newblock Society for Industrial and Applied Mathematics (2003)

\bibitem{saad1986}
Saad, Y., Schultz, M.H.: {GMRES}: {A} generalized minimal residual algorithm
  for solving nonsymmetric linear systems.
\newblock SIAM Journal on Scientific and Statistical Computing \textbf{7}(3),
  856--869 (1986).
\newblock \doi{10.1137/0907058}

\bibitem{salmietal2014}
Salmi, S., Toivanen, J., von Sydow, L.: An {IMEX}-scheme for pricing options
  under stochastic volatility models with jumps.
\newblock SIAM Journal on Scientific Computing \textbf{36}(5), B817--B834
  (2014).
\newblock \doi{10.1137/130924905}

\bibitem{scott2006}
Scott, B.: Gyrokinetic study of the edge shear layer.
\newblock Plasma Physics and Controlled Fusion \textbf{48}(5A), A387 (2006)

\bibitem{taitanoetal2016}
Taitano, W., Chac{\'o}n, L., Simakov, A.: An adaptive, conservative 0{D}--2{V}
  multispecies {R}osenbluth--{F}okker--{P}lanck solver for arbitrarily
  disparate mass and temperature regimes.
\newblock Journal of Computational Physics \textbf{318}, 391--420 (2016).
\newblock \doi{10.1016/j.jcp.2016.03.071}

\bibitem{taitanoetal2015}
Taitano, W., Chac{\'o}n, L., Simakov, A., Molvig, K.: A mass, momentum, and
  energy conserving, fully implicit, scalable algorithm for the
  multi-dimensional, multi-species {R}osenbluth--{F}okker--{P}lanck equation.
\newblock Journal of Computational Physics \textbf{297}, 357--380 (2015).
\newblock \doi{10.1016/j.jcp.2015.05.025}

\bibitem{thomasetal2012}
Thomas, A., Tzoufras, M., Robinson, A., Kingham, R., Ridgers, C., Sherlock, M.,
  Bell, A.: A review of {V}lasov--{F}okker--{P}lanck numerical modeling of
  inertial confinement fusion plasma.
\newblock Journal of Computational Physics \textbf{231}(3), 1051--1079 (2012).
\newblock \doi{10.1016/j.jcp.2011.09.028}.
\newblock Special Issue: Computational Plasma Physics

\bibitem{thomasetal2009}
Thomas, A.G.R., Kingham, R.J., Ridgers, C.P.: Rapid self-magnetization of laser
  speckles in plasmas by nonlinear anisotropic instability.
\newblock New Journal of Physics \textbf{11}(3), 033,001 (2009)

\bibitem{wolkeknoth2000}
Wolke, R., Knoth, O.: Implicit--explicit {R}unge--{K}utta methods applied to
  atmospheric chemistry-transport modelling.
\newblock Environmental Modelling \& Software \textbf{15}(6--7), 711--719
  (2000).
\newblock \doi{10.1016/S1364-8152(00)00034-7}

\bibitem{xuetal2010}
Xu, X., Bodi, K., Cohen, R., Krasheninnikov, S., Rognlien, T.: {TEMPEST}
  simulations of the plasma transport in a single-null tokamak geometry.
\newblock Nuclear Fusion \textbf{50}(6), 064,003 (2010)

\bibitem{xuetal2007}
Xu, X., Xiong, Z., Dorr, M., Hittinger, J., Bodi, K., Candy, J., Cohen, B.,
  Cohen, R., Colella, P., Kerbel, G., Krasheninnikov, S., Nevins, W., Qin, H.,
  Rognlien, T., Snyder, P., Umansky, M.: Edge gyrokinetic theory and continuum
  simulations.
\newblock Nuclear Fusion \textbf{47}(8), 809 (2007)

\bibitem{xuetal2008}
Xu, X.Q., Xiong, Z., Gao, Z., Nevins, W.M., McKee, G.R.: {TEMPEST} simulations
  of collisionless damping of the geodesic-acoustic mode in edge-plasma
  pedestals.
\newblock Phys. Rev. Lett. \textbf{100}, 215,001 (2008).
\newblock \doi{10.1103/PhysRevLett.100.215001}

\end{thebibliography}

\end{document}